%% file: hc_ewpt_v2.tex
\documentclass[12pt,superscriptaddress]{article}

\pdfoutput=1
\usepackage{jheppub}
\usepackage{graphicx}  
\usepackage{dcolumn}   
\usepackage{bm}        
\usepackage{amssymb}   
\usepackage{slashed} 

\newcommand{\beq}{\begin{equation}}
\newcommand{\eeq}[1]{\label{#1}\end{equation}}
\def\beqa{\begin{eqnarray}}
\def\eeqa#1{\label{#1}\end{eqnarray}}
\newcommand{\eeqn}{\end{equation}}
\newcommand{\CR}{\notag \\}
\newcommand{\leqn}[1]{(\ref{#1})}
\newcommand{\cO}{{\cal O}}                 
\newcommand{\cL}{{\cal L}}

\def\Veff{V_{\rm eff}}

\def\stacksymbols #1#2#3#4{\def\theguybelow{#2}
    \def\vp{\lower#3pt}
    \def\sp{\baselineskip0pt\lineskip#4pt}
    \mathrel{\mathpalette\intermediary#1}}

\def\intermediary#1#2{\vp\vbox{\sp
     \everycr={}\tabskip0pt
     \halign{$\mathsurround0pt#1\hfil##\hfil$\crcr#2\crcr
              \theguybelow\crcr}}}

\def\gsim{\stacksymbols{>}{\sim}{2.5}{.2}}
\def\lsim{\stacksymbols{<}{\sim}{2.5}{.2}}
\begin{document}

\title{Higgs Couplings and Electroweak Phase Transition}

\author[a]{Andrey Katz} 
\author[b]{and Maxim Perelstein}

\affiliation[a]{Department of Physics, Harvard University, Cambridge, MA 02138, USA}

\affiliation[b]{Laboratory for Elementary Particle Physics, 
	     Cornell University, Ithaca, NY 14853, USA}
	     
\emailAdd{andrey@physics.harvard.edu}     
\emailAdd{mp325@cornell.edu}     

\abstract{We argue that extensions of the Standard Model (SM) with a strongly first-order electroweak phase transition generically predict significant deviations of the Higgs couplings to gluons, photons, and $Z$ bosons from their SM values. Precise experimental measurements of the Higgs couplings at the LHC and at the proposed next-generation facilities will allow for a robust test of the phase transition dynamics. To illustrate this point, in this paper we focus on the scenario in which loops of a new scalar field are responsible for the first-order phase transition, and study a selection of benchmark models with various SM gauge quantum numbers of the new scalar. 
We find that the current LHC measurement of the Higgs coupling to gluons already excludes the possibility of a first-order phase transition induced by a scalar in a sextet, or larger, representation of the $SU(3)_c$. Future LHC experiments (including HL-LHC) will be able to definitively probe the case when the new scalar is a color triplet. If the new scalar is not colored, an electron-positron Higgs factory, such as the proposed ILC or TLEP, would be required to test the nature of the phase transition. The extremely precise measurement of the Higgsstrahlung cross section possible at such machines will allow for a comprehensive and definitive probe of the possibility of a first-order electroweak phase transition in all models we considered, including the case when the new scalar is a pure gauge singlet.}


\maketitle


\section{Introduction}

The recent discovery of a Higgs boson, with a mass of approximately 125 GeV, opens a new era of direct probes of electroweak symmetry breaking. Currently, the LHC data is consistent with the single Standard Model (SM) Higgs interpretation, with several rates measured at $20-30$\% level. 
In the coming years, much more precise measurements of the Higgs properties will be performed at the LHC, and, hopefully, at the next-generation lepton collider. Studies indicate that a per-cent level precision on many of the Higgs couplings can be realistically achieved~\cite{Dawson:2013bba}. It is therefore timely to consider physical implications of such high-precision Higgs measurements.  

While today we clearly live in a state with broken electroweak symmetry, it is expected that the symmetry is restored at sufficiently high temperatures, {\it e.g.} in the early Universe. A transition from the high-temperature symmetric phase to the low-temperature, broken-symmetry phase occurred about a nanosecond after the Big Bang. The dynamics of this transition is an open question, with potentially important implications. For example, a first-order phase transition, with significant entropy production, is required in scenarios of electroweak baryogenesis~\cite{Cohen:1993nk,Riotto:1999yt,Morrissey:2012db}, and may also produce potentially observable gravitational waves~\cite{Grojean:2006bp}. Theoretically, dynamics of the phase transition is determined by the structure of the Higgs effective potential (free energy) at finite temperature. While this object is not directly measurable at colliders, it is tightly connected to the properties of the Higgs boson at zero-temperature. One may therefore hope to gain useful information about the phase transition from precision Higgs data from collider experiments.    

If physics up to the TeV scale is completely described by the SM, it is well known that the electroweak phase transition (EWPT) 
is second-order~\cite{Arnold:1992rz,Kajantie:1996qd}. Although no direct experimental evidence for physics beyond the Standard Model (BSM) currently exists, theoretical arguments strongly suggest that such physics should exist. If so, the dynamics of the EWPT is model-dependent. 

Probably the best known scenario where the EWPT could be strongly first-order is the Minimal Supersymmetric Standard Model (MSSM) with a light 
stop. The stop has an appreciable coupling to the Higgs field, thus altering its effective potential and allowing a first-order 
EWPT~\cite{Huet:1995sh,Carena:1996wj,Carena:2008vj}. The Higgs mass and rates measured at the LHC strongly disfavor 
this possibility~\cite{Curtin:2012aa,Cohen:2012zza}, although some scenarios may still be possible, {\it e.g.} models with invisible Higgs decays into light neutralinos~\cite{Carena:2012np}. Regardless of the fate of the MSSM, 
one can easily imagine other models where a first-order EWPT is possible. These fall into one of the two classes:

\begin{itemize}

\item New physics in loops: New particles couple to the Higgs boson, but do not affect its tree-level potential. Loop corrections, however, may be large enough to alter the nature of the phase transition.  

\item New physics at tree level: The tree-level Higgs potential may differ from the SM, either due to mixing with other scalars, or due to higher-dimension operators~\cite{Grojean:2004xa}. Both effects may affect the phase transition.

\end{itemize}

\noindent In this paper, we focus on the first class of models. We do not commit to any specific, complete BSM scenario. Instead, we study a representative sample of simple toy models in which a first-order EWPT is possible. Our toy models have a very simple BSM matter content, just one new scalar field, allowing for a clear illustration of the underlying physics. Since only states that are light ($\lsim 400$~GeV) and have significant couplings to the Higgs can affect the EWPT dynamics, our results will in fact apply to a broad range of realistic BSM theories. 

While direct searches for new physics at the Tevatron and the LHC place strong bounds on many BSM models, they {\it do not} preclude the possibility of BSM scalars in the $100-400$ GeV mass range that we consider. Depending on the decay channels, even colored scalars in this mass range may be allowed. For example, the strongest current bound on a color-triplet diquark, decaying to two jets, is placed by the Tevatron experiments and is about 100~GeV~\cite{Aaltonen:2013hya}. While the LHC experiments may be able to 
improve the bound in this particular case~\cite{Bai:2013xla}, many other possibilities will likely escape direct detection 
even 
with the full LHC data set. These include, for example, a colored state decaying to four jets, or a gauge-singlet scalar coupled only to the Higgs and too heavy to participate in Higgs decays.  
On the other hand, any scalar which has a strong effect on the EWPT dynamics should be expected to modify the Higgs production cross sections and/or decay branching ratios. The connection between EWPT and the observable Higgs properties is direct, generic, and robust. Therefore, unlike the highly model-dependent direct searches, precision measurements of the Higgs properties could provide a definitive answer to the question of whether a first-order EWPT in the early Universe is possible or not. The goal of this paper is to demonstrate that this is indeed the case, and identify the relevant observables and levels of precision needed to address this question. 

More concretely, we will consider a single scalar\footnote{It is well known that scalar loops induce a cubic term in the high-temperature effective potential, providing a straightforward mechanism for a first-order EWPT. Fermion loops do not generate such a term. Nevertheless, in some cases it is possible to generate a first-order EWPT via fermion loops~\cite{Carena:2004ha}; this scenario is outside the scope of this paper. For a recent analysis of $h\to \gamma\gamma$ coupling deviations in such a model, see Ref.~\cite{Davoudiasl:2012tu}.} $\Phi$, coupled to the Higgs via
\beq
V \propto \kappa |\Phi|^2 |H|^2~. 
\eeq{eq:coup}
While in the MSSM $\kappa$ would be related to gauge and/or Yukawa couplings, here we consider it to be a free parameter, constrained only by perturbativity requirements. Assuming that $\kappa \sim \cO(1)$ (we will 
show in Sec.~\ref{sec:results} that this is in fact a necessary condition for a first-order EWPT), we expect the following Higgs observables to be modified:
\begin{enumerate}
\item If $\Phi$ is colored, the coupling of the Higgs to gluons, and, therefore, Higgs gluon fusion production cross section at the LHC. As we will see, 
this is already a powerful observable: for example, it completely excludes a first-order EWPT induced by a color-sextet $\Phi$. For the case when $\Phi$ is a color triplet, 
all of the parameter space with a first-order EWPT will be probed at a $3\sigma$ level at the LHC-14 with a 3 ab$^{-1}$ 
data set (HL-LHC).
\item If $\Phi $ is charged under $U(1)_{EM}$, the coupling of the Higgs to photons, and therefore BR($h \to \gamma \gamma$), is modified. This is 
potentially a spectacular observable. However, we will find that in many cases, a first-order EWPT is compatible with shifts in BR($h \to \gamma \gamma$) that are too small to be observed at the LHC-14.
\item At one-loop level, the Higgs coupling to $Z$ bosons is modified. This effect is present independently of the quantum numbers 
of $\Phi$, since $\Phi$ necessarily renormalizes the Higgs wavefuniction~\cite{Englert:2013tya,Craig:2013xia}. While numerically small, this correction may in fact be accessible at future electron-positron Higgs factories such as the ILC~\cite{Baer:2013cma} 
or TLEP~\cite{Gomez-Ceballos:2013zzn}, which can measure the Higgsstrahlung ($e^+e^-\to Zh$) cross section with a sub-percent precision. 
\end{enumerate}  
The first two of these points have been already 
studied in~\cite{Curtin:2012aa,Cohen:2012zza,Carena:2012np,Huang:2012wn}, for the particular case 
of $\Phi$ with the quantum numbers of the MSSM stop (and often with an extra assumption, that it has an MSSM
quartic coupling). 
In this paper we extend this analysis to a broader range of BSM scenarios.\footnote{For earlier work along similar lines, see Ref.~\cite{Chung:2012vg}.} In general, we find that when $\Phi$ is colored, the $hgg$ coupling provides the most sensitive probe. In fact, models with $\Phi$ in $SU(3)_c$ representations larger than a 
triplet, with a first-order EWPT, are already ruled out by the LHC data. For non-colored $\Phi$, we find that the cross section 
$\sigma(e^+e^-\to Zh)$ can provide a robust and sensitive probe of the EWPT. (Another robust probe is the Higgs cubic self-coupling~\cite{Noble:2007kk}; however, experimental measurements of this coupling with required accuracy are very challenging.) We will show that some of the models we consider predict 
deviations 
large enough to be observed at the ILC, while the projected sensitivity of TLEP is sufficient to probe the entire parameter 
space with a first-order EWPT in all models under consideration. 

The paper is organized as follows. Section~\ref{sec:TF} discusses the general theoretical framework for understanding the EWPT and the non-SM contributions to Higgs couplings in the class of models we consider, as well as defines the benchmark models used in our study. Section~\ref{sec:anal} contains a very simplified, analytic treatment of the EWPT, illustrating the connection between a first-order EWPT and the Higgs couplings corrections. The main results of our analysis, obtained via numerical treatment of the EWPT, are presented in Section~\ref{sec:results}. Finally, we discuss our results and outline some open questions in Section~\ref{sec:disc}. Appendix~\ref{app:1} contains a collection of results useful in the effective thermal potential calculation.

\section{Theoretical Framework}
\label{sec:TF}

In this section, we will outline the theoretical framework of our analysis, and present a general argument connecting the EWPT dynamics with the zero-temperature couplings of the Higgs boson. 

\subsection{Higgs Potential and Electroweak Phase Transition: The SM and Beyond}

In this paper, we assume that electroweak symmetry breaking is due to a single SM Higgs doublet $H$, with a 
tree-level potential given by
\beq
V_0 = -\mu^2|H|^2 + \lambda|H|^4.
\eeq{Vtree}
The measured Higgs vacuum expectation value (vev) $v=\mu/\sqrt{\lambda}=246$~GeV and Higgs boson mass $m_h=\sqrt{2}\mu = 126$~GeV determine the coefficients of this potential: 
\beq
\mu\approx 90~{\rm GeV}, \ \lambda\approx 0.13.
\eeq{values}
We assume that the dominant BSM correction to Higgs physics comes from loops of a single non-SM scalar field 
$\Phi$, whose tree-level contribution to the scalar potential is of the form
\beq
V_\Phi = m_0^2 |\Phi|^2 +  \kappa |\Phi|^2 |H|^2 + \eta |\Phi|^4.
\eeq{Vphi} 
We do not fix the SM gauge quantum numbers of  $\Phi$ at this point; we will consider several possibilities as described in 
Sec.~\ref{sec:models}. 

To study the EWPT dynamics, consider the effective finite-temperature potential $\Veff(\varphi; T)$, where $T$ is temperature. Physically, this object is just the free energy of the field configuration with a constant, spatially homogeneous Higgs field 
\beq
H_{\rm bg}=(0, \frac{\varphi}{\sqrt{2}})~,
\eeq{eq:higgs}
and all other fields set to zero. Including one-loop quantum corrections, the effective potential has the form
\beq
\Veff(\varphi; T) \,=\, V_0(H_{\rm bg}) + V_1(\varphi) + V_T(\varphi; T)\,,
\eeq{Veff}
where $V_1$ is the one-loop contribution to the zero-temperature effective potential (also known as Coleman-Weinberg potential), and 
$V_T$ is the thermal correction~\cite{Dolan:1973qd,Weinberg:1974hy}. Both $V_1$ and $V_T$ receive contributions from all particles coupled to the Higgs. A particle's contribution to both $V_1$ and $V_T$ is determined by its multiplicity $g_i$, its fermion number $F_i$, and its mass in the presence of a background Higgs field (or {\it Higgs-dependent mass} for short), $m_i(\varphi)$:
\beq
V_1(\varphi) = \frac{g_i (-1)^{F_i}}{64\pi^2} \left[ m_i^4(\varphi) \log \frac{m_i^2(\varphi)}{m_i^2(v)} - \frac{3}{2} m_i^4(\varphi) + 2 m_i^2(\varphi)m_i^2(v)\right];
\eeq{oneloopT0}
\beq
V_T(\varphi; T) = \frac{g_i T^4 (-1)^{F_i}}{2\pi^2} \int_0^\infty dx \, x^2 \log \left[ 1 - (-1)^{F_i} \exp \left( \sqrt{x^2 + \frac{m_i^2(\varphi)}{T^2}} \right) \right]\,,
\eeq{oneloopT} 
where $v=246$ GeV is the zero-temperature Higgs vev. Notice that $V_1$ includes the counterterms required to maintain the tree-level values of $\mu$ and $\lambda$ in Eq.~\leqn{values}. The multiplicity factors are normalized so that a gauge-singlet real scalar corresponds to $g=1$, while a gauge-singlet Dirac fermion gives $g=4$. In our analysis, we include the contributions of the BSM scalar $\Phi$, as well as the SM top quark and the electroweak gauge bosons; for details, see Appendix~\ref{app:1}. We ignore loops of other SM particles due to their small couplings to the Higgs.   

It is well known that thermal perturbation theory contains infrared divergences in the limit of zero boson mass, resulting in an enhancement of certain class of multi-loop diagrams, so-called ``daisy" diagrams, at large $T$. Fortunately, such diagrams can be resummed~\cite{Fendley:1987ef,Carrington:1991hz}. The resulting ``ring-improved" thermal potential is given by simply replacing $m_i^2(\varphi)$ in Eq.~\leqn{oneloopT} with the thermal mass:
\beq
m_i^2(\varphi) \rightarrow m_i^2(\varphi) + \Pi_i(T)\,,
\eeq{mT}
where $\Pi_i$ are the one-loop two-point functions at finite temperature. At large $T$, they can be approximated as $\Pi_i \approx c_i T^2$. The coefficients $c_i$ in the SM are listed in Appendix~\ref{app:1}, while the BSM contributions are summarized in Table~\ref{tab:models}. 

At high temperature, the thermal effective potential can be expanded as $V_T(\varphi, T) \sim A T^2 \varphi^2$, where $A$ depends on the particle content and couplings of the theory. In almost all known models, and certainly in all models studied here, $A>0$, meaning that the full effective potential has a minimum at $\varphi=0$. This minimum describes a state with unbroken electroweak symmetry, and at very high temperatures immediately after the Big Bang the Universe is in this state. (Here we make the standard and mild assumption that reheating temperature is well above the weak scale.) As the Universe cools, it transitions into the state of broken electroweak symmetry. In a first-order phase transition, the effective potential develops a local electroweak-symmetry breaking (EWSB) minimum at $\varphi_{\rm EWSB}\not=0$ while $\varphi=0$ is still the global minimum. Eventually, the EWSB minimum becomes energetically preferred, and the Universe tunnels into that state. We define the critical temperature $T_c$ to be the temperature at which the two vacua are degenerate. (Strictly speaking, the transition occurs at a somewhat lower temperature, but this difference is typically small.) The ``strength" of the transition can be characterized by a dimensionless ratio
\beq
\xi = \frac{\varphi_{\rm EWSB}(T_c)}{T_c}.
\eeq{xi}
Larger values of $\xi$ correspond to stronger deviations from quasi-adiabatic evolution, {\it i.e.} higher entropy production. Numerical studies show that a rough condition for successful electroweak baryogenesis is $\xi\gsim 0.9$. We will use this value as a rough boundary between the regions of parameter space with and without a strongly first-order transition. 

\subsection{Higgs Couplings}

In the class of models we consider, deviations from SM Higgs couplings are due to loops of $\Phi$ particles. An obvious place to 
look for such deviations is in couplings which first appear at the one-loop level in the SM, namely $hgg$ and $h\gamma\gamma$. 
Somewhat more surprisingly, we find that measurements of the $hZZ$ coupling can also play an important role in constraining the EWPT dynamics. Even though in this case the BSM loops appear as small corrections to the SM tree-level coupling, the very high precision with which this coupling can be measured in the Higgsstrahlung process at $e^+e^-$ Higgs factories makes it a sensitive probe of new physics. This probe is especially important in models where $\Phi$ is an SM gauge singlet, since in this case $hgg$ and $h\gamma\gamma$ couplings remain unaffected.

\subsubsection{Couplings to photons and gluons}

 The contribution of a particle with mass $\gg m_h$ to these couplings can be described by effective operators,
\beq
{\cal L}_{h\gamma \gamma} \,=\, \frac{2 \alpha}{9 \pi v} C_\gamma h F_{\mu \nu} F^{\mu \nu}\,,~~~
{\cal L}_{hgg} \,=\, \frac{\alpha_s}{12 \pi v} C_g h G_{\mu \nu} G^{\mu \nu}\,,
\eeq{Higgs_ops}  
where the normalization is chosen such that $C_\gamma=C_g=1$ for the SM top quark at one loop. Here $h$ is the physical Higgs boson, $H = (H^+, \frac{v+h}{\sqrt{2}})$, and $F_{\mu\nu}$ and $G_{\mu\nu}$ are the $U(1)_Y$ and $SU(3)_c$ field strength tensors, respectively. The contributions of a new heavy scalar $\Phi$ can be found using the well-known ``low-energy theorems"~\cite{Ellis:1975ap, Shifman:1979eb}. At one loop, these contributions are given by
\beq
C^\Phi_g = \frac{1}{4} C(r_\Phi) \frac{\partial \ln m_\Phi^2(\varphi)}{\partial \ln \varphi}\,,~~~C^\Phi_\gamma=\frac{3}{64}g_{\Phi}Q_\Phi^2 \frac{\partial \ln m_\Phi^2(\varphi)}{\partial \ln \varphi}~,
\eeq{Phi_h}
where 
$Q_\Phi$ is the electric charge of $\Phi$, and the coefficient $C(r)$ is defined by Tr$(t_a^r t_b^r)=C(r) \delta_{ab}$ . The fractional deviations of the $hgg$ and $h\gamma\gamma$ decay amplitudes from the SM are
\beq
R_g\equiv \frac{{\cal A}(hgg)}{{\cal A}(hgg)|_{\rm SM}}\,=\,C_g,~~~R_\gamma\equiv \frac{{\cal A}(h\gamma\gamma)}{{\cal A}(h\gamma\gamma)|_{\rm SM}}\,\approx\,1-0.27 \left(C_\gamma-1\right)\,,
\eeq{Rdefs}
where the contribution of the $W$ loop has been taken into account in the photon coupling. 
Notice that the non-SM contributions to the Higgs couplings are determined by {\it exactly the same object}, the Higgs-dependent mass of the field $\Phi$, as the effective Higgs potential, see Eqs.~\leqn{oneloopT0} and~\leqn{oneloopT}. For this reason, one should expect a generic, robust connection between the coupling shifts and the EWPT dynamics.\footnote{This argument is very similar to the one made in~\cite{Farina:2013ssa} to establish a similarly robust connection between the shifts in these couplings and naturalness of electroweak symmetry breaking.}  In particular, large deviations from the SM in the 
effective potential, required for a strongly first-order EWPT, should correspond to large, observable corrections to SM Higgs couplings. 
In the rest of this paper, we will quantify this connection.

\subsubsection{Coupling to $Z$s}

An exception to the above argument occurs when the $\Phi$ field is neither colored nor electrically charged. Such a field can still drive a first-order EWPT, if it is strongly coupled to the Higgs and/or has a large multiplicity factor, {\it e.g.} due to a BSM global symmetry~\cite{Espinosa:2007qk}. It obviously does not contribute (at one-loop) to $hgg$ or $h\gamma\gamma$ couplings. However, it does induce a one-loop contribution to the Higgs wavefunction renormalization. Experimentally, the best place to search for this effect is in the $e^+e^-\to hZ$ cross section, which can be measured with a very high precision at a next-generation electron-positron collider. 
If the $\Phi$ field is an SM gauge singlet, the fractional deviation of this cross section from its SM value is given by~\cite{Englert:2013tya,Craig:2013xia}
\beq
\delta_{hZ} = - \frac{g_\Phi \kappa^2 v^2}{24 \pi^2 m_h^2} \left( 1 + F(\tau_\Phi)\right)\,,
\eeq{hZshift}
where $\tau_\Phi = m_h^2/(4m_\Phi^2)$, and 
\beq
F(\tau_\Phi) = \frac{1}{2\sqrt{\tau_\Phi(1-\tau_\Phi)}} 
\arctan \left[ \frac{2\sqrt{\tau_\Phi(1-\tau_\Phi)}}{2\tau_\Phi-1}\right]\,.
\eeq{Fform}
For small $\tau_\Phi$, $F(\tau_\Phi)=-1-\frac{2}{3}\tau_\Phi+\ldots$, 
so that the shift in $\delta_{hZ}$ decouples in the large $m_\Phi$ limit. 

Below, we will also apply Eq.~\leqn{hZshift} to models in which $\Phi$ is {\it not} an SM gauge singlet, and thus has direct gauge couplings to the $Z$. In those models, the one-loop contribution to the $e^+e^-\to hZ$ cross section contains the vertex correction 
and the $Z$ wavefunction renormalization pieces as well. However, those corrections are subdominant to the Higgs wavefunction renormalization, as noted in Ref.~\cite{Englert:2013tya}. One reason for this is that the Higgs wavefunction is the only correction which scales as $\kappa^2$, the others scaling as $\kappa g^2$ and $g^2$; in our case, $\kappa\gg g^2$ throughout the interesting parameter region.

It was shown in~\cite{Craig:2013xia} that this deviation can be used as a powerful probe of naturalness in models where the 
top loop quadratic divergence in the Higgs mass parameter is canceled by a non-colored partner 
({\it e.g.}, ``folded SUSY''~\cite{Burdman:2006tz}). 
Typically, these models predict an
$\cO(1\%) $ deviation from the SM value, which should be observable either at TLEP or at the ILC. However, the effect 
is much more general: any new particle with significant coupling to the Higgs will inevitably contribute. We will show in Sec.~\ref{sec:results} that the entire parameter 
space where the first-order EWPT is driven by an SM gauge-singlet $\Phi$ can be probed at TLEP. Moreover, we will show that even in some cases where $\Phi$ is electrically charged, this deviation can be easier to probe than deviation in the $h \gamma \gamma$ coupling, given the projected experimental sensitivities in the two channels at $e^+e^-$ Higgs factories. 

\subsection{Benchmark Models}
\label{sec:models}

	\begin{table}[h]
\begin{center}
\begin{tabular}{|l|c|c|c|c|c|c|c|}
			\hline
			Model~~~~  &   $(SU(3), SU(2))_{U(1)}$    &  $g_\Phi$ & $C_3$ &  $C_2$ & $\frac{\Pi_W}{g^2T^2}$  & $\frac{\Pi_B}{g^{\prime 2}T^2}$ & $\frac{\Delta \Pi_h}{\kappa T^2}$  \\
          \hline
            ``RH stop" & $(\bar{3}, 1)_{-2/3}$ & $6$ & $4/3$ & 0 & $11/6$ & $107/54$ & $1/4$  \\
            Exotic triplet & $(3, 1)_{-4/3}$ & $6$ & $4/3$ & 0 & $11/6$ & $131/54$ & $1/4$  \\
            Exotic sextet & $(\bar{6}, 1)_{8/3}$ & $12$ & $10/3$ & 0 & $11/6$ & $227/54$ & $1/2$  \\
	  ``LH stau" & $(1, 2)_{-1/2}$ & $4$& 0 & $3/4$ & $2$ & $23/12 $ & $1/6$  \\
	  ``RH stau" & $(1, 1)_{1}$ & $2$& 0 & 0 & $11/6 $ & $13/6 $ & $1/12$  \\
	  Singlet & $(1, 1)_0$ & $2$ & 0 & 0 & $11/6$ & $11/6$ & $1/12$  \\
	\hline
			\end{tabular}
\caption{Benchmark models studied in this paper.}
\label{tab:models}
\end{center}
	\end{table}

To illustrate the connection between EWPT dynamics and Higgs couplings, we will study several benchmark models, which differ in the SM gauge quantum numbers assigned to the BSM scalar field $\Phi$. The models are summarized in Table~\ref{tab:models}. Note that we label some of the models with 
the names of a SUSY particle with quantum numbers of $\Phi$, the right-handed stop and left-handed/right-handed stau; however, in these cases as in all others, the coupling constants $\kappa$ and $\eta$ are unconstrained. For each model, in addition to the quantum numbers of $\Phi$, we list its multiplicity $g_\Phi$, its $SU(3)$ and $SU(2)$ quadratic Casimirs $C_3(r)$ and $C_2(r)$, as well as the thermal masses of the SM gauge and Higgs bosons in the high-temperature limit. The thermal masses of the gauge bosons, $\Pi_W$ and $\Pi_B$, include both the SM and the $\Phi$ loop contributions. For the Higgs, we list only the additional contribution due to $\Phi$ loops; the SM contributions are discussed in Appendix~\ref{app:1}. The thermal mass of the $\Phi$ itself is given by
\beq
\frac{\Pi_\Phi}{T^2} = \frac{g^2 C_2(r)}{4} + \frac{g_s^2 C_3(r)}{4} +  
\frac{g^{\prime 2} Y_\Phi^2}{4} + \frac{\kappa}{6} +
 \frac{\eta}{6}\left( \frac{g_\Phi}{2} + 1\right)\,,
\eeq{phi_Pi}
where $g_s$, $g$ and $g^\prime$ are the SM $SU(3)_c$, $SU(2)_L$ and $U(1)_Y$ gauge couplings, respectively.

\subsubsection{Direct Collider Constraints on the Benchmark Models}
\label{sec:direct}

In this paper we will mostly consider BSM scalars in a physical mass range $\sim 100 \ldots 400$~GeV, some of them colored. 
One might naively expect that most of them are already excluded by direct Tevatron and LHC searches. In this short subsection 
we show that it is not the case, and many viable scenarios are still essentially unconstrained by direct searches. Moreover, many
of them will be very hard to constrain directly, and, therefore, the Higgs couplings that we exploit in this paper are going 
to provide \emph{the only robust handle} which will allow us either to discover, or to exclude these particles. 

Let us start with the colored particles. First, it is almost impossible to discuss the direct searches in a
completely model independent way, and we should specify possible decay modes. The first two benchmark models in Table~\ref{tab:models} can be perfect examples of 
``diquarks'', namely particles which are pair-produced, and each of which decays into a pair of jets. 
As was shown in~\cite{Giudice:2011ak}, these particles are safe from the point of view of FCNCs, while direct 
searches only constrain their mass to be $m \gsim 100$~GeV~\cite{Aaltonen:2013hya}. Therefore we conclude that if these
are indeed diquarks, they are unconstrained in the relevant mass range. Of course, the $\Phi$ in the ``RH stop" benchmark model could also be a ``true stop'' of R-parity conserving SUSY 
(while other superpartners are heavy, and their impact on the phenomenology can be safely neglected), and in this case
it is mostly excluded in the interesting mass range~\cite{Aad:2012xqa, Aad:2012uu, Chatrchyan:2013xna}, 
except for a small ``island'' of stealth stops. However, we should bear in mind that these strong bounds are only applicable to a particular decay mode, $t$+MET, and  
do not constrain, for example, diquarks with the quantum numbers of RH stop.\footnote{For a discussion of collider constraints
on very light stops in the context of R-parity conserving scenarios, and open possibilities in this context, 
see also~\cite{Krizka:2012ah}.} 

Our third benchmark model has gigantic production cross sections for a new colored scalar (since it is a sextet of $SU(3)$), 
and if it had been a diquark, it would have been excluded by straightforward diquark LHC searches, 
see e.g.~\cite{Aad:2011yh, ATLAS:2012ds, Chatrchyan:2013izb}. However, $\Phi\sim(\bar{6}, 1)_{8/3}$ cannot be coupled to the 
SM fields through a renormalizable operator. The lowest order coupling we can write down is $\cL \propto \Phi (u^c)^4$, 
which a-priori implies a complex decay pattern, potentially including secondary vertices, tops and multiple jets. We 
are not aware of any direct LHC search which might exclude such a particle in general. However, as we will see in Sec~\ref{sec:results}, 
it is in fact excluded simply by Higgs production rates in gluon fusion. 

Our last three benchmark models are even more evasive, since in these models $\Phi$ is uncolored and has a very small 
production cross section at the LHC.
Particles with such small cross sections can probably be discovered only if they have spectacular 
decay modes ({\it e.g.} all-leptonic decay, 
not including $\tau$), and in general can be considered unconstrained above the LEP bounds, generically 
$m \gsim 100$~GeV.\footnote{For example, particles from benchmark points~4 and~5 have quantum numbers of $\tilde \tau$
in SUSY, and therefore, can mostly decay into $\Phi \to \tau \tilde \chi^0$, yielding a signature of two taus in the final state 
(assuming pair-production) and MET. This signature is extremely difficult and to the best of our knowledge no meaningful 
bound has been put on this scenario by the LHC. Of course, this is not the only possibility, and other options are also possible, 
e.g. when a doublet $\Phi$ decays into two jets through $\Phi Q d^c$ coupling. } 
Clearly, the last option (the SM singlet) is not even produced directly, and therefore it is hard to imagine that it is can be found 
in a hadron collider, unless through modification of the SM Higgs decay modes (or by introducing new rare exotic decay modes, 
e.g. $h \to$~invisible). Therefore, we conclude that generally all our benchmark models are unconstrained by current direct searches in the $\Phi$ mass region relevant for our analysis.    
    
\section{EWPT/Higgs Coupling Connection: Analytic Treatment}
\label{sec:anal}

Before presenting numerical results, let us consider a much-simplified treatment of the problem which can be carried through analytically. Even though the 
approximations made here are often not strictly valid in examples of real interest, this analysis nevertheless provides a qualitatively correct and useful illustration of the physics involved. 

To drive a first-order EWPT, the BSM scalar $\Phi$ should provide the dominant loop contribution to the Higgs thermal potential at $T\sim T_c$. Let us therefore ignore the SM contributions. If $T_c$ is significantly higher than all other mass scales in the problem, a high-temperature expansion of the thermal potential can be used to analyze the phase transition, and zero-temperature loop corrections to the effective potential can be ignored. For simplicity, we will also omit the resummed daisy graph contributions to the thermal potential. In this approximation,
\beq
V_T(\varphi; T) \approx \frac{g_\Phi m_\Phi^2(\varphi) T^2}{24} - \frac{g_\Phi m_\Phi^3(\varphi) T}{12\pi} + \ldots
\eeq{highT}
The $\Phi$ mass in the presence of a background Higgs field is given by
\beq
m^2_\Phi(\varphi) = m_0^2 + \frac{\kappa}{2} \varphi^2.
\eeq{mPhi}  
If $m_0$ is sufficiently small, the second term in the thermal potential~\leqn{highT} is effectively cubic in $\varphi$. Such a negative $\varphi^3$ term can result in a stable EWSB minimum of the potential at high temperature, as required for first-order EWPT. Motivated by this, let us consider the case $m_0=0$, which allows for simple analytic treatment. The effective potential is 
\beq
\Veff(\varphi; T) = V_0(\varphi) + V_T(\varphi; T) \approx \frac{1}{2} \left(-\mu^2+ \frac{g_\Phi \kappa T^2}{24} \right) \varphi^2 - \frac{g_\Phi \kappa^{3/2} T}{24\sqrt{2}\pi} \varphi^3 +  \frac{\lambda}{4} \varphi^4.
\eeq{Veffm0}
The unbroken symmetry point $\varphi=0$ is a local minimum as long as 
\beq
\frac{g_\Phi \kappa T^2}{24} - \mu^2  > 0.
\eeq{v0good}
The location of the other minimum is given by the larger root, $\varphi_+$, of the quadratic equation
\beq
\lambda \varphi^2 -\frac{g_\Phi \kappa^{3/2} T}{8\sqrt{2}\pi} \varphi -\mu^2+ \frac{g_\Phi \kappa T^2}{24} = 0.
\eeq{vEW}
The critical temperature $T_c$ for the first-order transition is determined by the condition
\beq
V(0; T_c) = V(\varphi_+(T_c); T_c).
\eeq{Tcanal}
Solving Eqs.~\leqn{vEW},~\leqn{Tcanal} yields
\beq
T_c^2 = \frac{24\mu^2}{g_\Phi \kappa\left( 1-\frac{g_\Phi\kappa^2}{24\pi^2\lambda}\right)},~~~\varphi_+(T_c) = \frac{g_\Phi \kappa^{3/2}T_c}{12\sqrt{2}\pi\lambda}.
\eeq{solutions}
Requiring that a first-order transition occurs, $T_c^2>0$, and is strongly first-order, $\varphi_+(T_c)/T_c>1$, yields a range of acceptable values of $\kappa$:
\beq
\frac{5.5}{g^{1/2}_\Phi} \,> \kappa \,> \frac{3.6}{g^{2/3}_\Phi}. 
\eeq{kappas} 
As an example, consider a color-triplet, weak-singlet $\Phi$ field, as in the ``RH stop" or ``Exotic Triplet" benchmark models of Table~\ref{tab:models}. In this case, our estimate suggests that a strongly first-order transition occurs for values of $\kappa$ between 1.1 and 2.2. At the same time, the $\Phi$ loop contribution to the Higgs-gluon coupling is
\beq
R_g = \frac{1}{8} \ \frac{\kappa v^2}{m_0^2 + \frac{\kappa v^2}{2}}\,.
\eeq{hgg_0} 
In the limit $m_0^2\ll \frac{\kappa v^2}{2}$, which for $\kappa\sim 1$ corresponds to a broad range of $m_0$, we obtain $R_g \approx 1/4$, or a 25\% enhancement in the $hgg$ coupling compared to the SM. In fact, even larger enhancements are possible for negative values of $m_0^2$. Of course, the $hgg$ deviations from the SM become small when $m_0^2\gg \frac{\kappa v^2}{2}$; however, in this regime, the $\Phi$ mass is well above the weak scale, and it does not affect the EWPT dynamics either. Thus, models with first-order EWPT should produce a large effect, of order 10\% or more, in the Higgs-gluon coupling. This conclusion will be confirmed by the numerical analysis in the following section.

\section{Results}
\label{sec:results}

We developed a numerical code to analyze the dynamics of the electroweak phase transition in each of the benchmark models listed in Table~\ref{tab:models}. Given the model and the values of the free parameters, $m_0$, $\kappa$ and $\eta$, the code computes the effective potential as a function of temperature, Eq.~\leqn{Veff}, and identifies the critical temperature $T_c$. The $x$ integral in the finite-temperature potential~\leqn{oneloopT} is performed numerically, with no high-temperature approximation. This is important since the critical temperature in our models is typically of order 100 GeV, which is at the same scale as the masses of the particles involved. To identify the region in the parameter space of a given model where a strongly first-order phase transition occurs, we compute $T_c$ and $\xi$ on a dense grid of points in this space. We then analyze the deviations of the Higgs couplings from the SM in this region.  

\begin{figure}[tb]
\begin{center}
\centerline {
\includegraphics[width=3.0in]{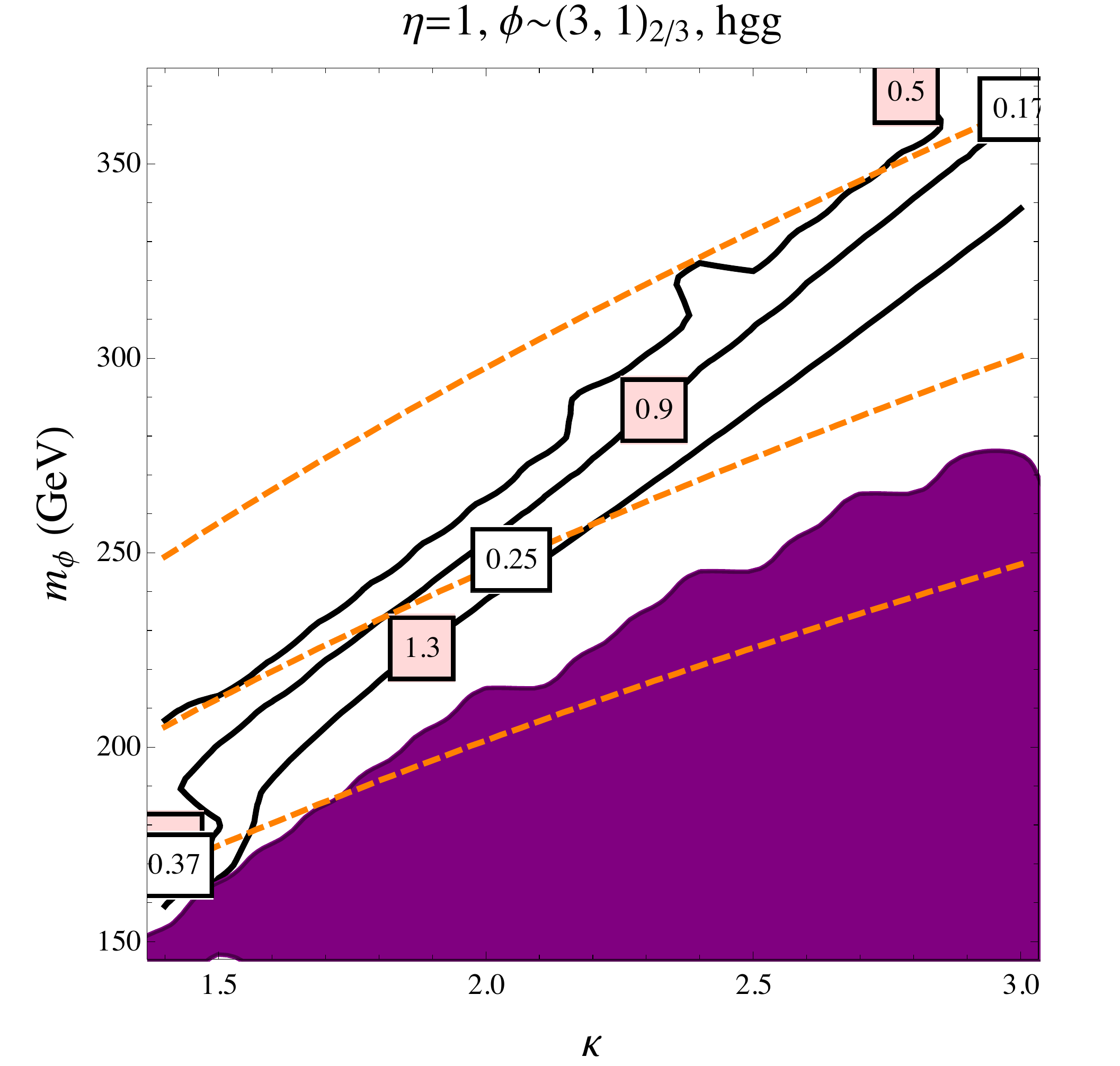}
\includegraphics[width=3.0in]{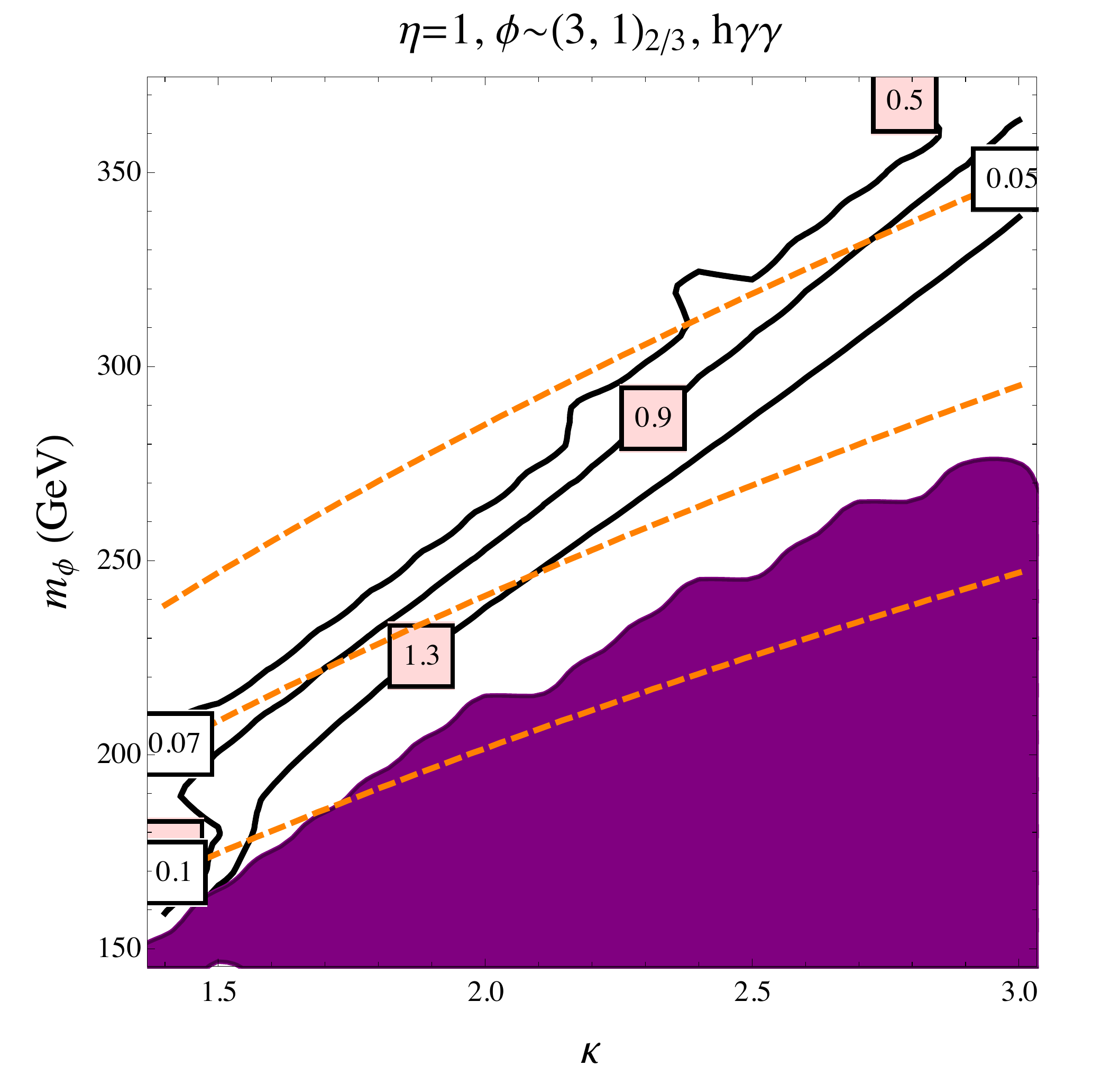}
}
\caption{The region of parameter space where a strongly first-order EWPT occurs in the ``RH stop" benchmark model. Also shown are the fractional deviations of the $hgg$ (left panel) and $h\gamma\gamma$ (right panel) couplings from their SM values. Solid/black lines: contours of constant EWPT strength parameter $\xi$ (see Eq.~\leqn{xi}). Dashed/orange lines: contours of constant $hgg/h\gamma\gamma$ corrections. (For the case of $h\gamma\gamma$ the correction is always negative, and the plots show its absolute value.) In the shaded region, phase transition into a color-breaking vacuum occurs before the EWPT.}
\label{fig:RHstop}
\end{center}
\end{figure}

\begin{figure}[tb]
\begin{center}
\centerline {
\includegraphics[width=3.0in]{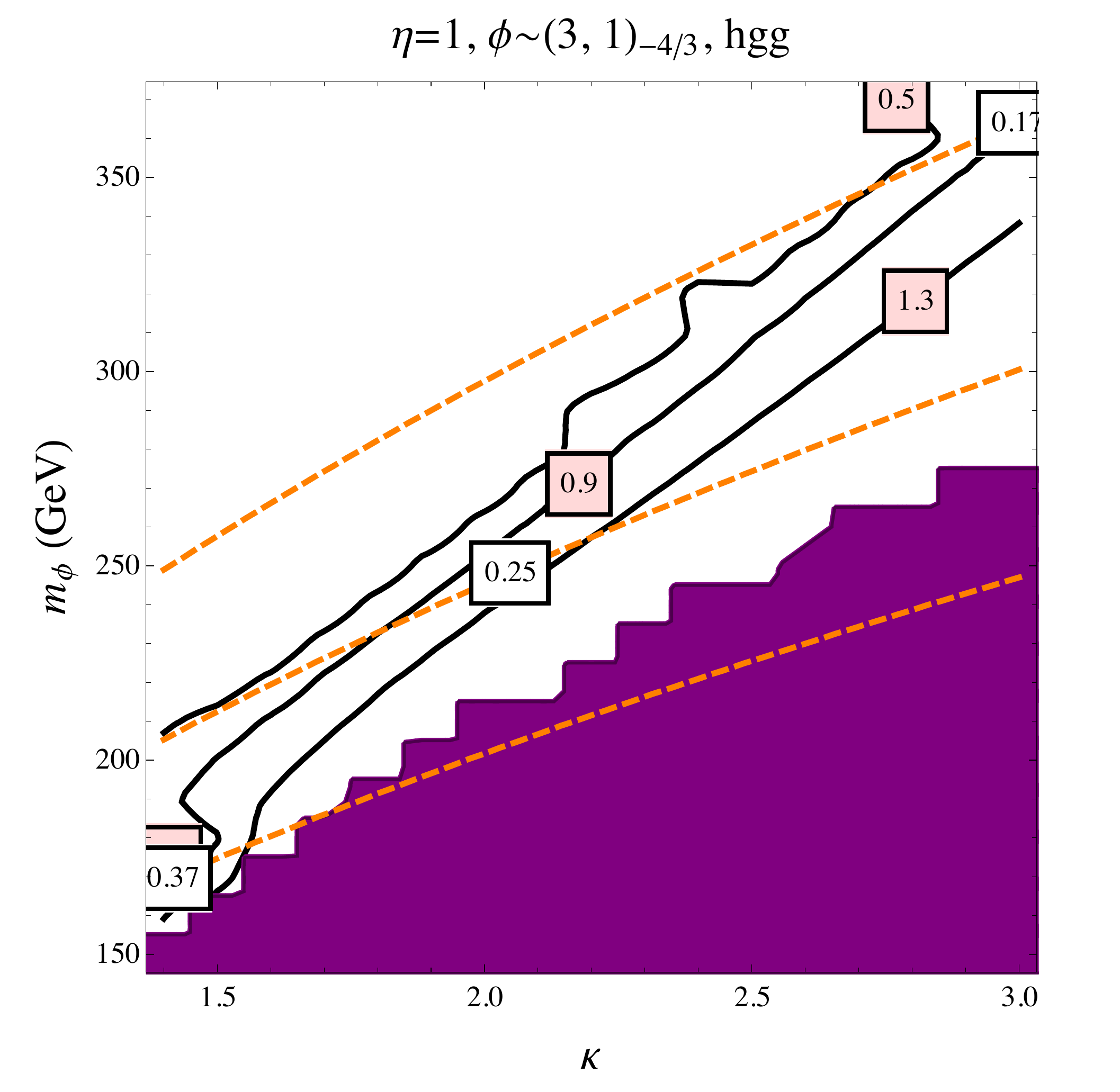}
\includegraphics[width=3.0in]{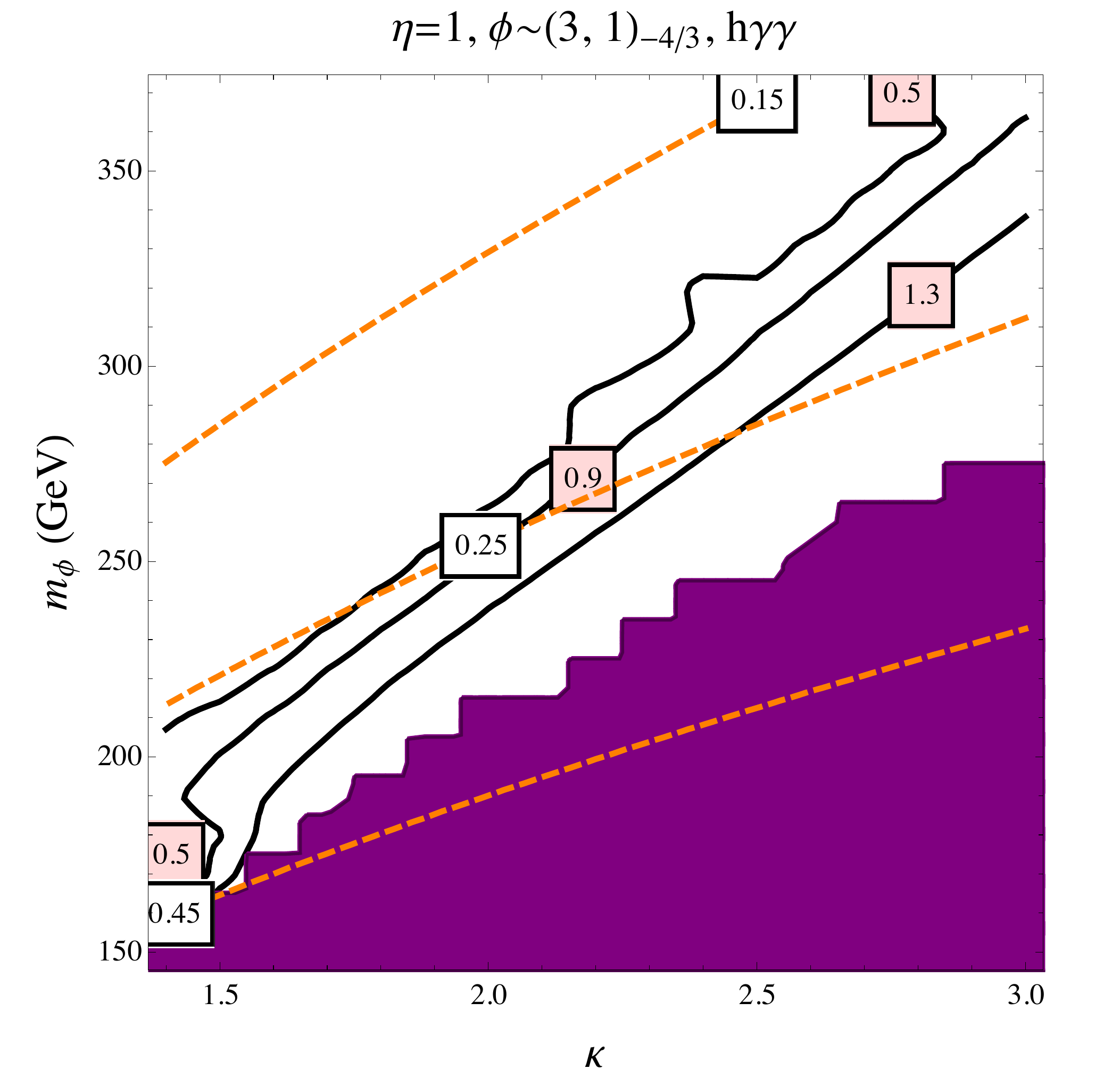}
}
\caption{Same as Fig.~\ref{fig:RHstop}, for the Exotic Triplet model (see Table~\ref{tab:models}).}
\label{fig:Diquark}
\end{center}
\end{figure}

\begin{figure}[tb]
\begin{center}
\centerline {
\includegraphics[width=3.0in]{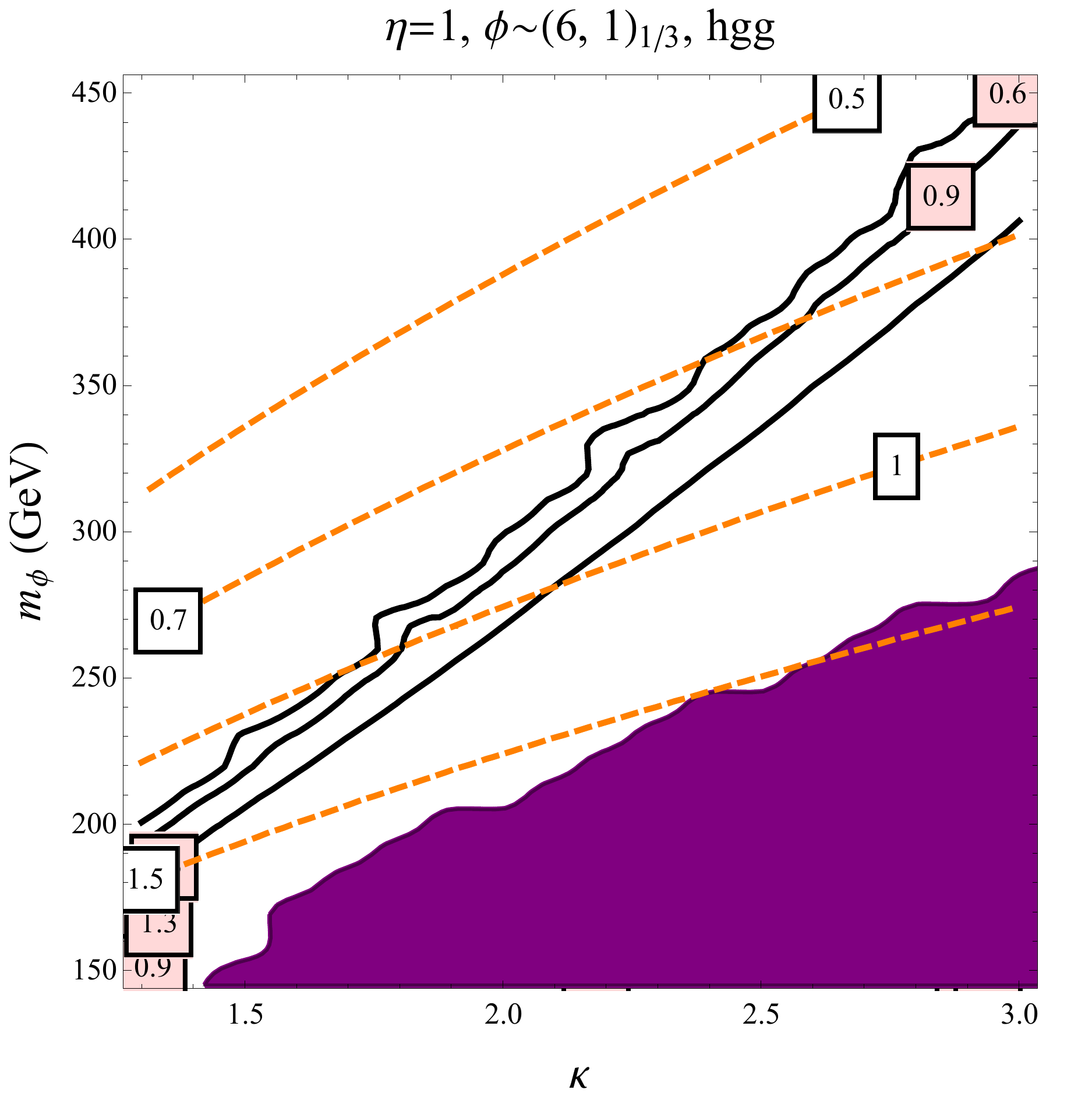}
\includegraphics[width=3.0in]{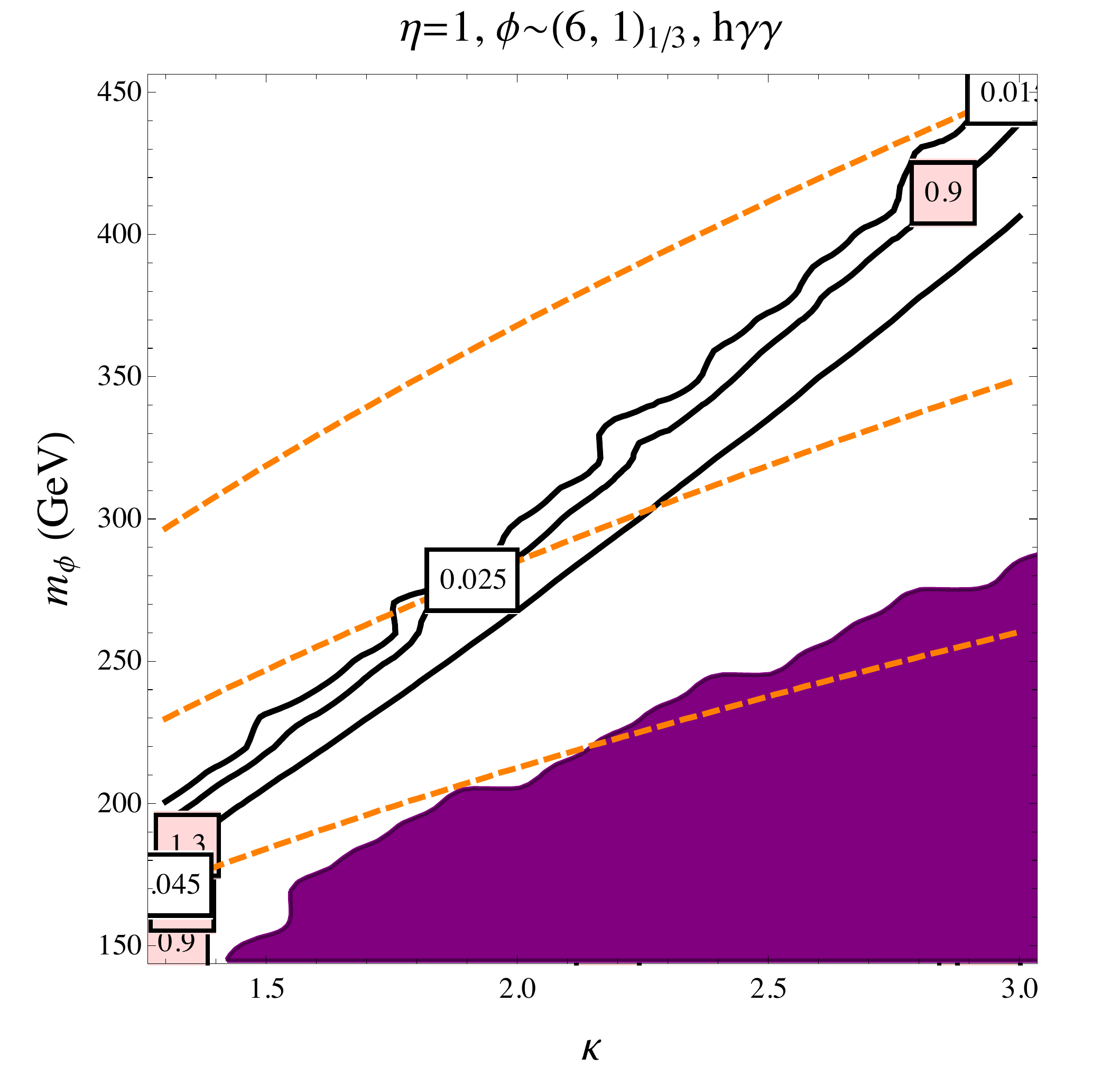}
}
\caption{Same as Fig.~\ref{fig:RHstop}, for the Sextet model (see Table~\ref{tab:models}).}
\label{fig:Sextet}
\end{center}
\end{figure}

\begin{figure}[tb]
\begin{center}
\centerline {
\includegraphics[width=3.0in]{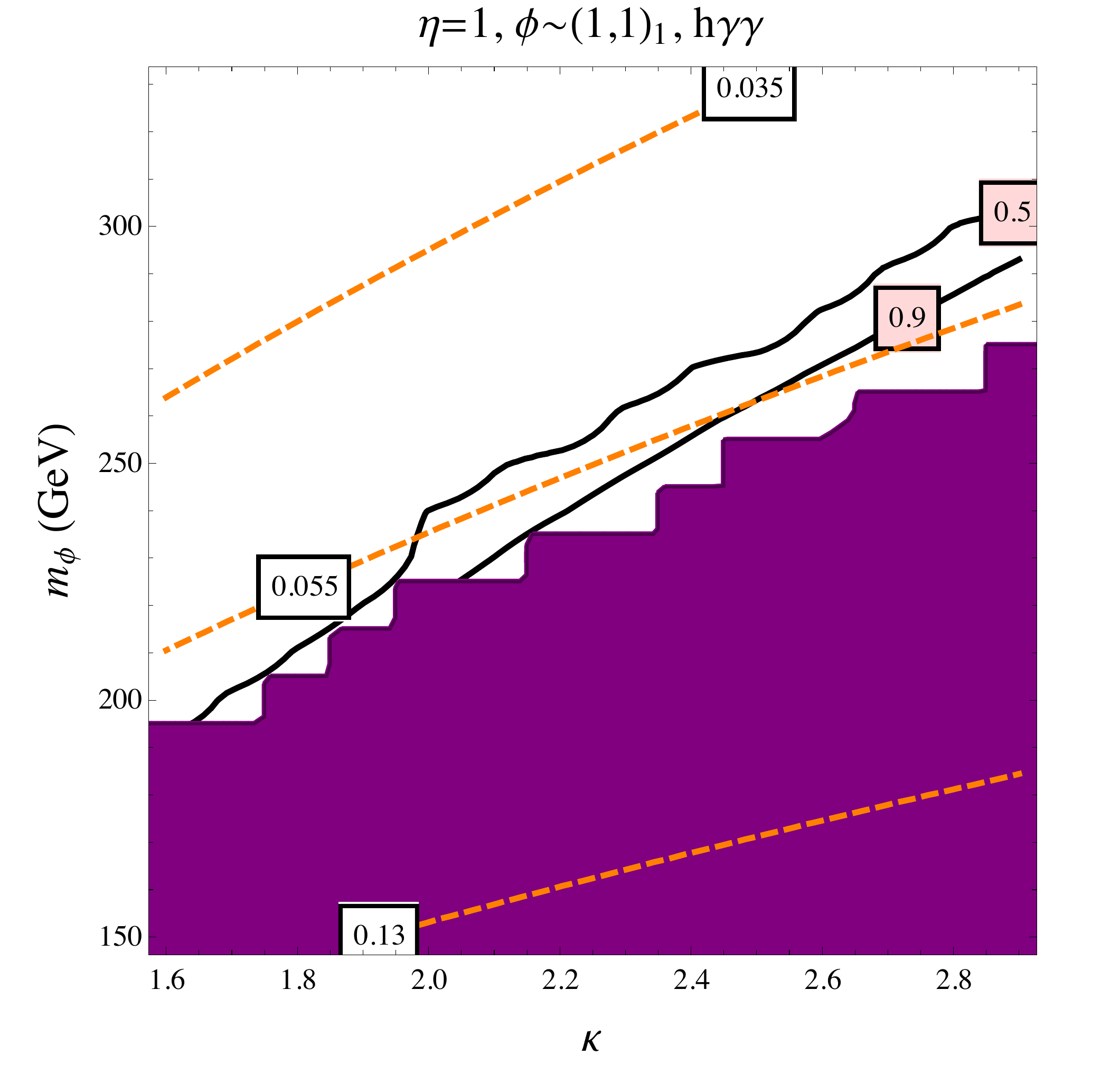}
\includegraphics[width=3.0in]{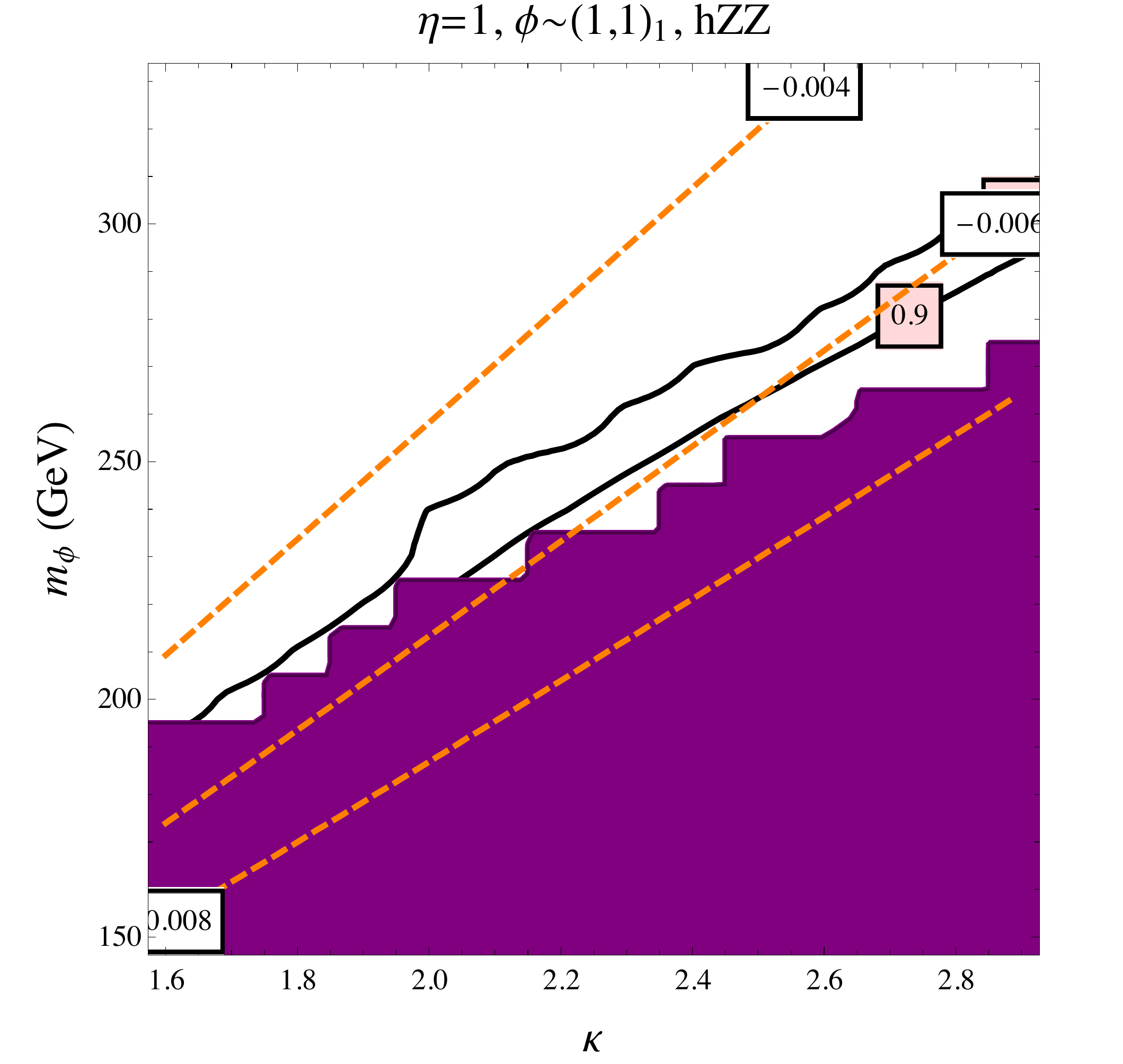}
}
\caption{The region of parameter space where a strongly first-order EWPT occurs in the ``RH stau" benchmark model. Also shown are the fractional deviations of the $h\gamma\gamma$ coupling (left panel) and the $e^+e^-\to hZ$ cross section (right panel) from their SM values. Solid/black lines: contours of constant EWPT strength parameter $\xi$ (see Eq.~\leqn{xi}). Dashed/orange lines: contours of constant $h\gamma\gamma/\sigma_{hZ}$ corrections. (The $h\gamma\gamma$ correction is always negative, and the plot shows its absolute value.) In the shaded region, phase transition into a wrong EM-breaking vacuum occurs before the EWPT.}
\label{fig:RHstau}
\end{center}
\end{figure}

\begin{figure}[tb]
\begin{center}
\centerline {
\includegraphics[width=3.0in]{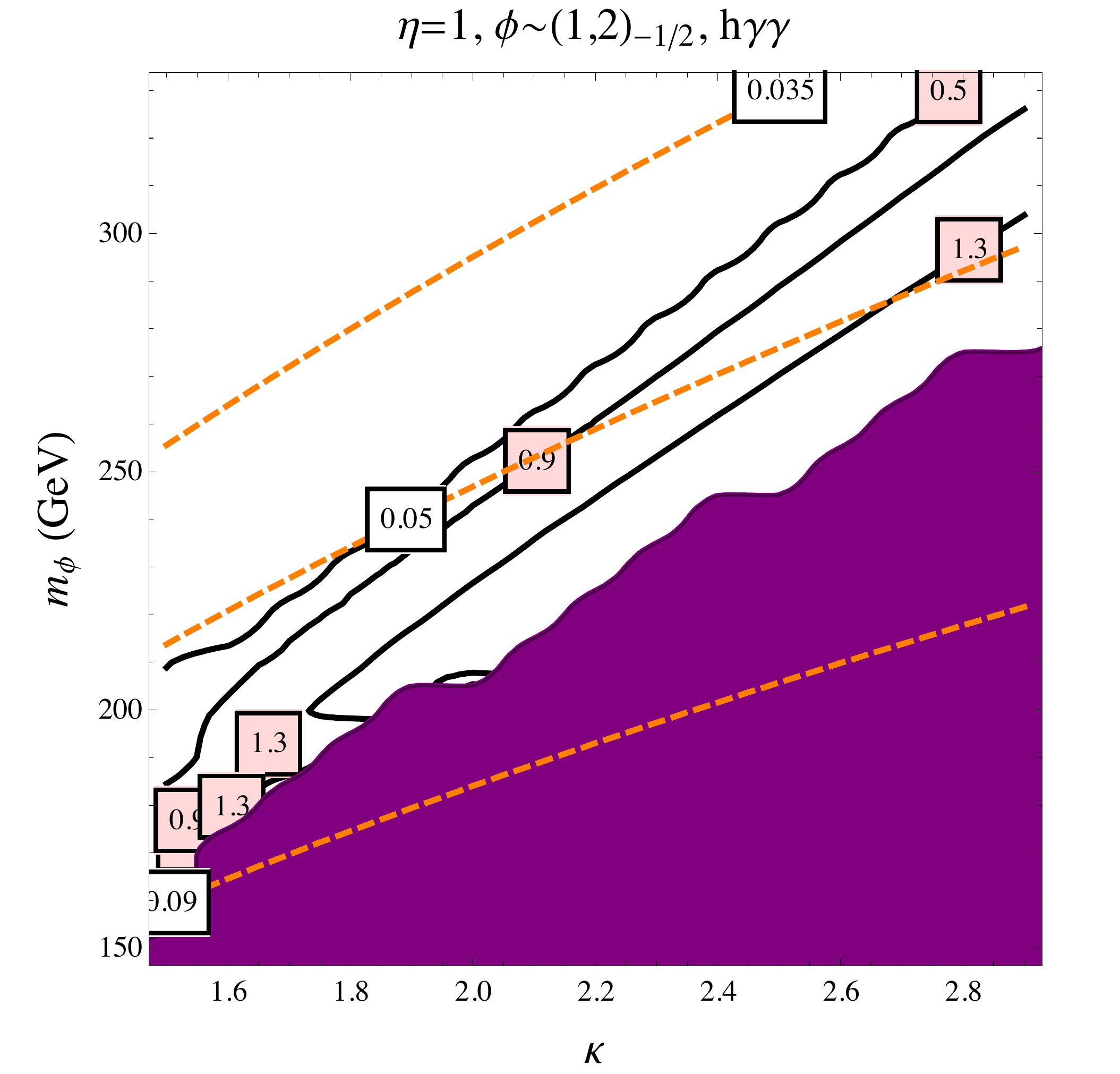}
\includegraphics[width=3.0in]{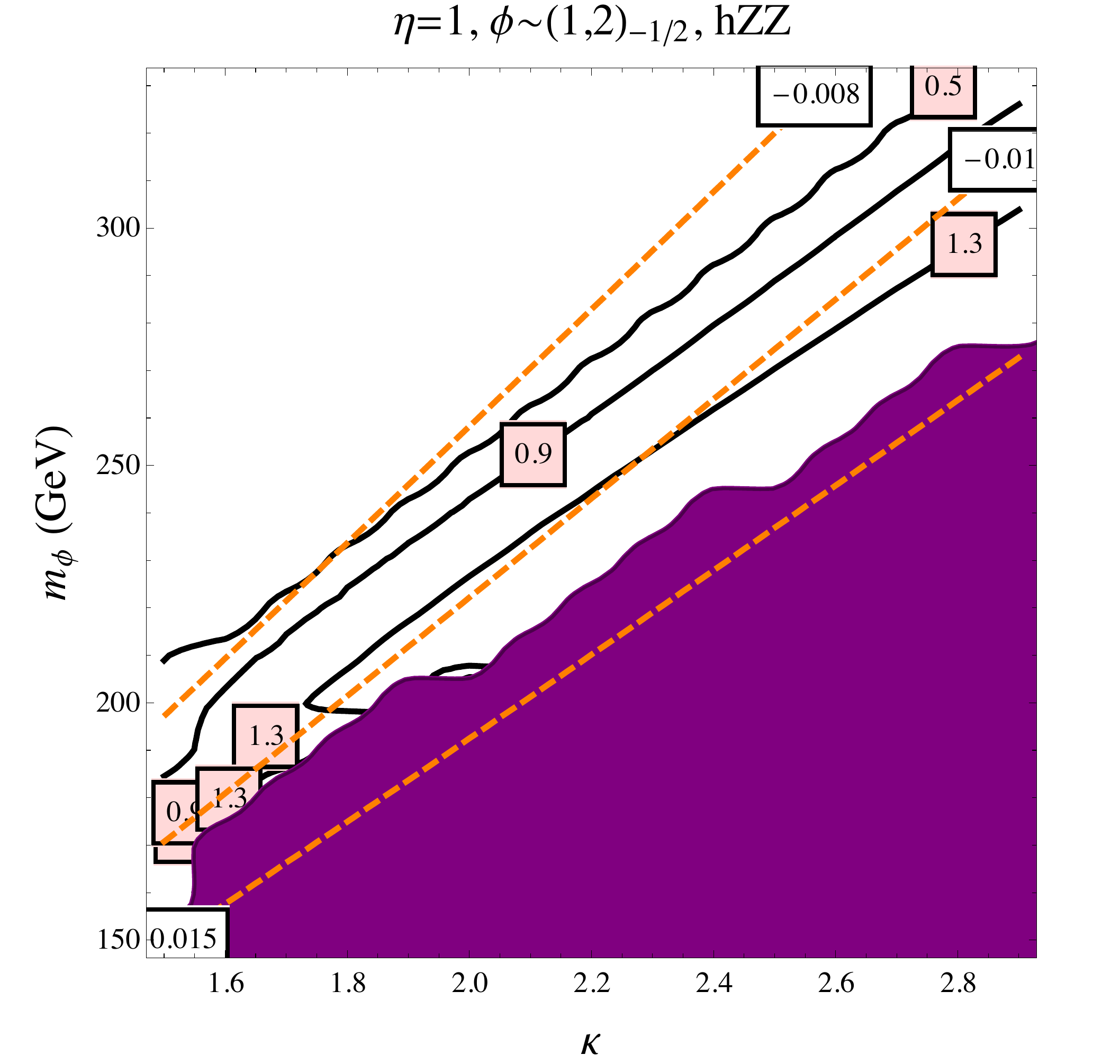}
}
\caption{Same as Fig.~\ref{fig:RHstop}, for the ``LH stau" model (see Table~\ref{tab:models}).}
\label{fig:LHstau}
\end{center}
\end{figure}

\begin{figure}[tb]
\begin{center}
\centerline {
\includegraphics[width=3.0in]{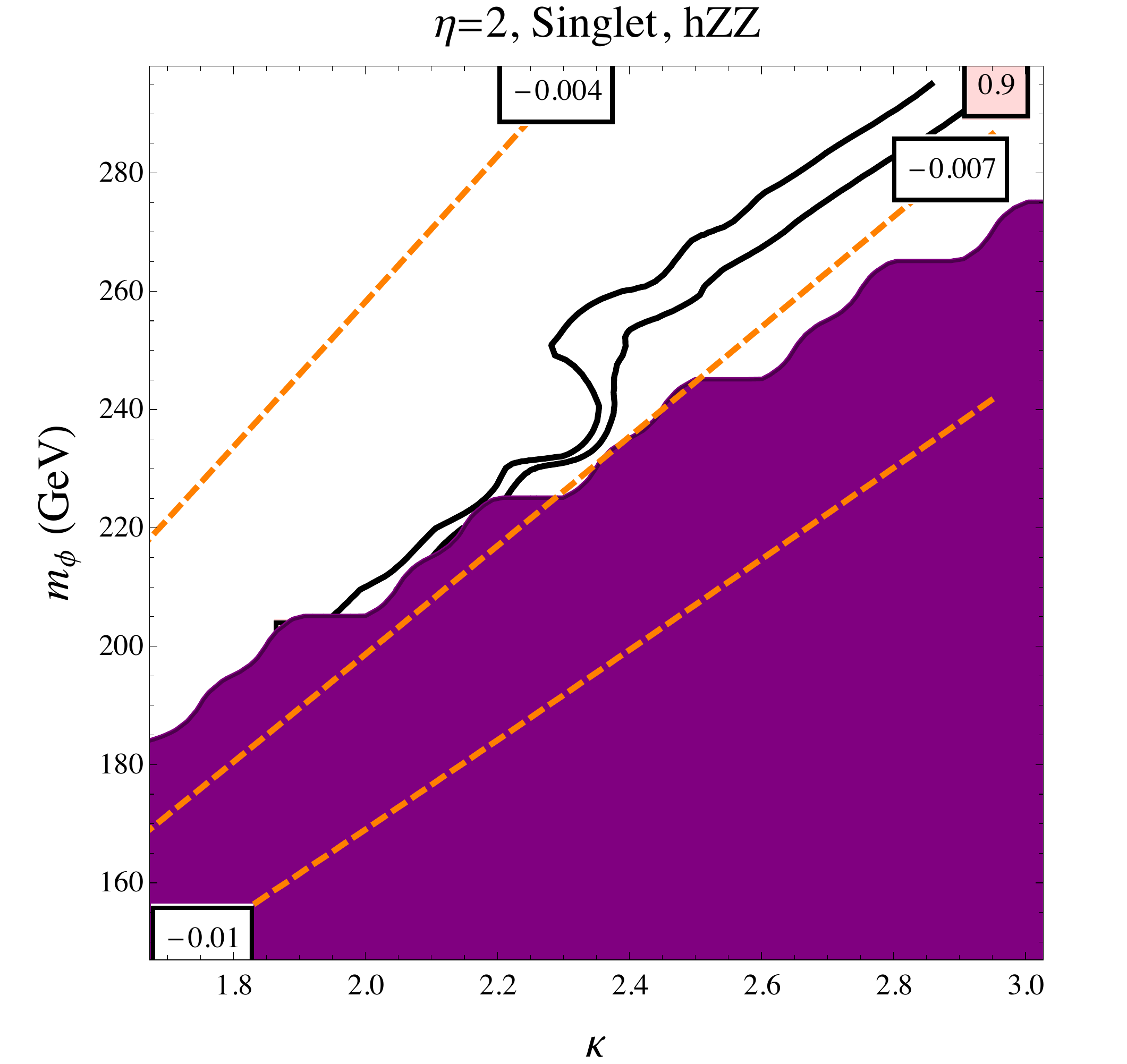}
\includegraphics[width=3.0in]{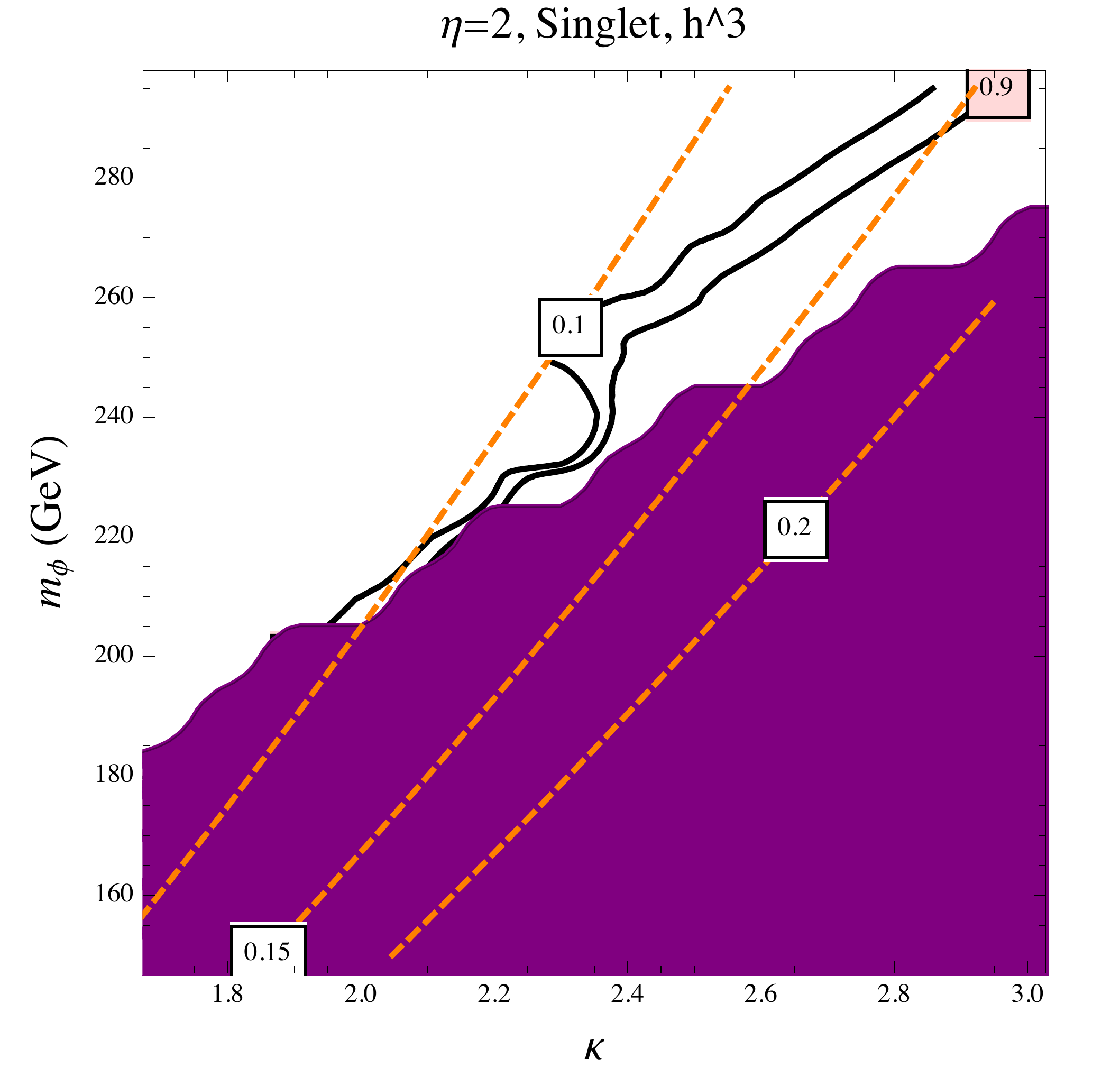}
}
\caption{The region of parameter space where a strongly first-order EWPT occurs in the Singlet benchmark model. Also shown are the fractional deviations of the $e^+e^-\to hZ$ cross section (left panel) and Higgs cubic self-coupling (right panel) from their SM values. Solid/black lines: contours of constant EWPT strength parameter $\xi$ (see Eq.~\leqn{xi}). Dashed/orange lines: contours of constant $\sigma_{hZ}/\lambda_3$ corrections. In the shaded region, phase transition into a wrong vacuum 
(with $\langle \phi \rangle \neq 0$) occurs before the EWPT.}
\label{fig:singlet}
\end{center}
\end{figure}

The results of this analysis are summarized in Figs.~\ref{fig:RHstop}--\ref{fig:singlet}. In all benchmark models, we fixed $\eta=1.0$, with the exception of the Singlet model in which this value of $\eta$ produces no viable parameter region with a first-order EWPT; in this case, we choose $\eta=2.0$. The coupling constant $\kappa$ is scanned between roughly 1.0 and 3.0; for smaller $\kappa$, no points with a strong first-order EWPT have been found, while for higher $\kappa$, perturbative expansion of the effective potentials is questionable. In the plots, we use the physical, zero-temperature mass of the $\Phi$ scalar, given by
\beq
m_{\rm phys}^2 = m_0^2 + \frac{\kappa v^2}{2}.
\eeq{mphys}
We scan $m_{\rm phys}$ between (roughly) 150 and 400 GeV; we do not find points with a strongly first-order EWPT (and perturbative $\kappa$) outside of this range. (Once again, we emphasize that such relatively light scalars are still allowed by direct searches, even if they are colored; see Sec.~\ref{sec:direct}.) Note that for some points in the scanned region, the ``bare" (pre-EWSB) mass$^2$ of the $\Phi$ field may be negative, $m_0^2<0$. In this case, it is possible that the system will undergo a phase transition in which $\Phi$ develops a vev, at a temperature {\it above} the $T_c$ found by our code. The shaded regions in the plots of this section indicate the parts of the parameter space satisfying the condition
\beq
m_0^2 + \Pi_\Phi(T_c)<0\,,
\eeq{shaded}
which implies that a phase transition into a ``wrong" (non-EWSB) vacuum takes place at some $T>T_c$. If this scenario occurs, our analysis of the phase transition dynamics is no longer valid, since it assumed that no fields other than $H$ get a vev. While we do not claim that the shaded regions are necessarily ruled out (for example, the Universe may undergo a second EWSB phase transition resulting in a phenomenologically acceptable vacuum at late 
times; see e.g. Ref.~\cite{Cline:1999wi} for a related discussion), 
the cosmological evolution in this case is much more complicated, and we will not consider it here. In any case, as will be clear from our plots, the deviations of the Higgs couplings in the shaded region are {\it larger} than in the regions we consider ``allowed"; therefore, the statements we will make concerning the minimal experimental precision required to conclusively probe the first-order EWPT scenarios in each model would still apply if portions of the shaded regions turn out to be phenomenologically acceptable. 

By the same token, we do not incorporate the constraint of stability (or metastability) of the standard EWSB vacuum at zero temperature, which also may play a role for negative $m_0^2$. In order to impose this constraint, one would have to analyze a full two-dimensional potential in $H$ and $\Phi$ directions. Such an analysis was performed in a model with a real scalar and a Higgs, in Ref.~\cite{Espinosa:2011ax}; it should be possible to generalize it to the case of complex scalar considered here, although such a study is outside the scope of our paper. We emphasize again that if some of the regions included in our plots turned out to be being ruled out by this constraint, this would only strengthen our conclusions. 

For the benchmark models with colored scalar (RH stop, Exotic Triplet and Sextet), we plot the contours of fractional deviation of the $hgg$ and $h\gamma\gamma$ couplings from their SM values. Note that the $hgg$ coupling is enhanced in all models we study, while the $h\gamma\gamma$ coupling is always suppressed. For comparison, the current bounds on these couplings reported by the ATLAS collaboration~\cite{ATLAS-CONF-2013-034} are
\beqa
R_g &=& 1.08 \pm 0.14,\CR
R_\gamma &=& 1.23^{+0.16}_{-0.13}.
\eeqa{ATLASbounds}
These results already have interesting implications for the possibility of a strongly first-order EWPT. In particular, the Sextet model, where the deviations in the $hgg$ coupling in the region with first-order EWPT are predicted to be about 60\% or above, is completely excluded.\footnote{A potential loophole that should be kept in mind is that these bounds assume no sizable BSM contributions to the Higgs width. If such a contribution is allowed, a 60\% deviation in the $hgg$ coupling is only excluded at a 2 sigma level, and thus the Sextet model remains marginally compatible with data.} It is clear that models where $\Phi$ is in even larger representations of $SU(3)_c$, {\it e.g.} an octet, are also ruled out. The RH Stop and Exotic Triplet models, on the other hand, are still compatible with data at 68\%~CL. However, a dramatic improvement in precision expected in future experiments will allow these models to be probed. In both models, the {\it minimal} deviation in the $hgg$ coupling compatible with a strongly first-order EWPT is about 17\%. A recent Snowmass study~\cite{Dawson:2013bba} estimated that this coupling can be measured with a precision of  $6-8$\% at the LHC-14, $3-5$\% at HL-LHC, 2\% at the ILC, 1\% at the ILC with a luminosity upgrade, and 0.8\% at TLEP. (Note also that while the LHC numbers make certain assumptions about the total width, the $e^+e^-$ machines can measure the $hZZ$ coupling without such assumptions, establishing a firm model-independent normalization for all measurements.) If no deviations from the SM are seen in the $hgg$ coupling after such precise measurements, the possibility of a first-order EWPT driven by a single colored scalar will be conclusively ruled out. We find that for models with colored BSM scalars, the $h\gamma\gamma$ measurement is not as sensitive as $hgg$: the projected sensitivities for the two couplings are similar, but the predicted size of the effect in the photon coupling is smaller due to the large SM $W$-loop contribution to this coupling.

In models where the BSM scalar is not colored, the $hgg$ coupling remains at its SM value. The LH Stau and RH Stau models provide examples where the BSM scalar is electrically charged, and modifies the $h\gamma\gamma$ coupling. The minimal shift in this coupling compatible with a strongly first-order EWPT is about $4-5$\% in both models. This is clearly too small to be constrained by the present data, but may be probed by future experiments. The Snowmass study~\cite{Dawson:2013bba} projects a precision of about 2\% at an upgraded ILC running at $\sqrt{s}=1$ TeV, and about 1.5\% at TLEP, enabling the entire region of parameter space with a first-order EWPT to be probed at a $\sim 3$ sigma level. Interestingly, a precise measurement of the Higgsstrahlung cross section at a future $e^+e^-$ Higgs factory could provide an even more sensitive probe in these models. The minimal shift in this cross section compatible with a first-order EWPT is about 0.8\% in the LH Stau model, and 0.6\% in the RH Stau model. The projected precision at ILC-500 (with a luminosity upgrade) is about 0.25\%, while TLEP is projected to measure this cross section with an impressive $0.05$\% accuracy. Such a measurement would provide a definitive probe of the possibility of a first-order EWPT in these models. 

Finally, if the BSM scalar responsible for the first-order EWPT is neither colored nor electrically charged, electron-positron Higgs factories can still explore this scenario by measuring the $e^+e^-\to hZ$ cross section, and the Higgs cubic self-coupling. This is illustrated in Fig.~\ref{fig:singlet}. The minimal fractional deviation in the $hZ$ cross section compatible with a first-order EWPT is about $0.6$\%, similar to the ``stau" models above. This can be probed at a $\sim 2.5$ sigma level at an upgraded ILC-500, and comprehensively tested at TLEP. In contrast, the predicted deviations in the Higgs cubic self-coupling are in the $10-20$\% range, making them difficult to test at the proposed facilities. (The accuracy of the self-coupling measurement at an ILC-1T with luminosity upgrade is estimated to be about 13\%~\cite{Dawson:2013bba}, while at TLEP it can be measured with a precision of about 30\% via its contribution to Higgsstrahlung~\cite{McCullough:2013rea}.) Thus, it appears that the Higgsstrahlung cross section provides the most sensitive probe of this challenging scenario.

\section{Discussion}
\label{sec:disc}

In this paper, we considered several toy models which can induce a first-order electroweak phase transition in the early Universe. In {\it all} models, we found a strong correlation between the strength of the phase transition and the deviations of the Higgs couplings from the SM. This suggests that precise measurements of the Higgs couplings have a potential to definitively determine the order of the electroweak phase transition. Such a determination would be not only fascinating in its own right, but would also have implications for other important questions in particle physics and cosmology, such as viability of electroweak baryogenesis.

We emphasize that an electron-positron Higgs factory, such as the proposed ILC or TLEP, plays an {\it absolutely crucial role} in determining the order of the phase transition. Models where the BSM scalar responsible for a first-order EWPT is colored can be probed at the LHC, with HL-LHC providing a coverage of the relevant parameter space at $>3$ sigma level in all such models. However, scenarios where the first-order EWPT is due to a {\it non-colored} BSM scalars are just as plausible. LHC will not be able to probe these scenarios: in fact, even when $\Phi$ is electrically charged, the shift it induces in $h\to\gamma\gamma$ in the region compatible with a first-order EWPT is too small to be probed even at the HL-LHC. On the other hand, $e^+e^-$ Higgs factories will be able to comprehensively explore such scenarios, primarily due to a very precise measurement of the Higgsstrahlung cross section, $\sigma(e^+e^-\to Zh)$. The impressive sensitivity of this measurement expected at the ILC and, especially, at TLEP, makes it a uniquely robust and powerful tool for addressing the issue of EWPT dynamics. 

An important limitation of our analysis is that all our benchmark models have a {\it single} scalar field. The most important new effect in the presence of multiple fields with masses around the weak scale is the possibility of accidental cancellations in the BSM loop contributions to Higgs couplings. For example, in the MSSM, the stop sector contribution to Higgs coupling to gluons and photons is approximately given by~\cite{Arvanitaki:2011ck,Blum:2012ii}
\beq
C_g-1 = C_\gamma -1 = \frac{1}{4}\left( \frac{m_t^2}{m_{\tilde{t}_1}^2}  + \frac{m_t^2}{m_{\tilde{t}_2}^2} - \frac{m_t^2 X_t^2}{m_{\tilde{t}_1}^2m_{\tilde{t}_2}^2} \right),
\eeq{SUSY_Rg}
where $m_{\tilde{t}_i}$ are the stop eigenmasses, and $X_t=A_t-\mu/\tan\beta$. It is possible for the last term to cancel the first two, reducing or nullifying the deviations of these couplings from the SM. Since functional dependence of the effective thermal potential on the stop masses is quite different, the stop effects there would not cancel, 
and the possibility of a first-order transition may remain open even after very precise measurements of $R_g$ and $R_\gamma$. (Note that in the MSSM itself this possibility is not realized due to Higgs mass constraint; it would require a model where two light stop-like particles are compatible with a 125 GeV Higgs.) Of course, there is no known symmetry to enforce the cancellation in Eq.~\leqn{SUSY_Rg}, so such a scenario would require fine-tuning. More interestingly, it appears that this scenario should still be testable by a precise measurement of $\sigma(e^+e^-\to Zh)$ at an electron-positron Higgs factory. Since the contribution of each stop mass eigenstate to the Higgs wavefunction renormalization is proportional to the {\it square} of the stop-Higgs coupling, these contributions  should be additive, and thus should be of the same order as in our single-field models. We leave a detailed analysis of this interesting issue for future work.

Another potential issue for our analysis is the importance of higher-order corrections. For example, two-loop corrections to the thermal potential are known to be sizable in the SM~\cite{Arnold:1992rz} and the MSSM~\cite{Espinosa:1996qw,Carena:1997ki}. In general, we expect that two-loop QCD corrections could be important in benchmark models with colored $\Phi$. To partially address this issue, we compared the EWPT strength parameter $\xi$ for the ``RH stop" model, computed in the one-loop approximation of this paper, with the two-loop results of Ref.~\cite{Cohen:2011ap}. We found that our results are in good qualitative agreement with Fig.~2 of~\cite{Cohen:2011ap}, indicating that qualitative conclusions of our study should apply after two-loop corrections are taken into account. 

It is also well known that the thermal loop expansion for EW baryogenesis is borderline, since the 
thermal loop expansion parameter is 
${\cal O}(1)$. It is true that two-loop corrections might somewhat improve the precision of the calculation, but in order 
to get a fully trustworthy estimate, a full non-perturbative treatment is needed. Recent lattice studies~\cite{Laine:2012jy,Laine:2013raa}
show that perturbative calculations tend to slightly underestimate the strength of the EWPT, so that the parameter regions with a strongly first-order EWPT are in reality somewhat larger than suggested by our calculations. This would not affect the qualitative conclusions of our work, but in the future it would be very interesting to apply non-perturbative techniques to the sequence of toy models considered here to get a better estimate of the Higgs coupling measurement accuracy required for a complete probe of first-order EWPT.  

Finally, as stated in the Introduction, this paper only considered one of the two mechanisms for obtaining a first-order EWPT: new physics in loops. There are of course many models where a first-order EWPT is due to tree-level effects, such as mixing of the Higgs with other fields or higher-dimension operators. In such models, Higgs couplings are typically already modified at tree-level, which should lead to even larger deviations from the SM than in the cases considered here. A comprehensive study of the Higgs couplings/EWPT correlations in this class of models would be worthwhile.

\vskip0.8cm
\noindent{\large \bf Acknowledgments} 
\nopagebreak
\vskip0.3cm
\nopagebreak

We would like to thank Nima Arkani-Hamed, Nathaniel Craig, Tim Cohen, David Curtin, 
Matthew McCullough, Ann Nelson, and Matt Reece for useful discussions. We are also grateful to Yosef Nir for pointing 
us out several important typos in the first version of the paper. 
The authors are grateful to the Aspen physics center (winter), where this work was initiated. AK is supported by the Center for Fundamental Laws of Nature at Harvard. 
MP is supported by the U.S. National Science Foundation through grant PHY-0757868 and CAREER grant PHY-0844667. MP would like to acknowledge the hospitality of KITP at Santa Barbara, where part of this work was completed. 

\appendix

\section{Thermal Mass Formulas}
\label{app:1}

\input{appendix}

\bibliography{lit}
\bibliographystyle{jhep}
\end{document}

%% file: appendix.tex
In this Appendix we will review the calculation of the thermal masses in the SM, as well as in a generic BSM scenario. We will largely 
follow the calculations of~\cite{Comelli:1996vm} and for details the reader is referred to this paper and references therein.

We parametrized the Higgs field as follows: 
\beq
H = \frac{1}{\sqrt{2}} \left( \begin{array}{c}
\chi_1 + i \chi_2 \\
\varphi + h + i \chi_3
\end{array}
\right),
\eeq{eq:higgsparametrization}  
with $\varphi$ denoting the background field and $h$ the physical Higgs perturbation. As usual, we define $\chi^\pm=(\chi_1 \pm i \chi_2)/\sqrt{2}$. 
The SM contribution to the thermal masses of the $W^\pm$ gauge bosons is given by 
\beqa \nonumber
\Pi_{W_L^\pm} & = & \frac{g_2^2 T^2}{24} \Big( 3 (\theta_{W^\pm} + \theta_{W_3}) + 12 \theta_{W_3} \theta_{W^\pm} + 2N_c N_f \theta_{U_L} 
\theta_{D_L} + 2 N_f \theta_{\nu_L} \theta_{e_L} + (\theta_h + \theta_{\chi^\pm})^2  - 2 \theta_{\chi_3}\Big) \\  
& = & \frac{11}{6} g_2^2 T^2~,
\eeqa{eq:thermgaugemasswpm}
where $g_2$ is the $SU(2)_L$ gauge coupling, and $\theta_i \equiv \theta(T-m_i)$ is a step function equal to $1$ if $T > m$ and $0$ if $T < m$. 
The second line is the high-temperature approximation, valid for $T>m_t$. Any new $SU(2)$ doublet contributes to this thermal mass 
\beq
\Delta \Pi_{W_L^\pm}^{scalar,\ 2} = \frac{g_2^2 T^2}{24} \ N_c (\theta_u + \theta_d)^2, \ \ \ \ \Delta \Pi_{W_L^\pm}^{fermion,\ 2} = 
\frac{g_2^2 T^2}{24} \
2 N_c \theta_u \theta_d~,
\eeq{eq:scal2piwpm}
where the normalization in the second formula corresponds to a single Weyl fermion.
Note that here we assumed that different components of the $SU(2)$ doublet, which we denote by subscripts ``u'' and ``d'', may have different masses. 
Generalizing to arbitrary representations $r$ and neglecting the splittings between up- and down-component we get\footnote{For expressions with 
large splittings see Eqs.~(47) and~(50) in~\cite{Comelli:1996vm}.} 
\beq
\Delta \Pi_{W_L^\pm}^{fermion, r} = \frac{g_2^2 T^2}{6} {\rm Tr}\ T^+(r) T^-(r), \ \ \ \ 
\Delta \Pi_{W_L^\pm}^{scalar, r} = \frac{g_2^2 T^2}{6} 2 {\rm Tr}\ T^+(r) T^-(r)
\eeq{eq:wpmarb}
with 
\beq
T^\pm (r) \equiv \frac{T^1(r) \pm i T^2(r)}{\sqrt{2}}~.
\eeq{eq:tpm}

Similarly, the SM contribution to the thermal mass of $W^3$ is
\beqa \nonumber
\Pi_{W_L^3 } & = & \frac{g_2^2 T^2}{24} \Big( 18\theta_{W^\pm} +N_c N_f (\theta_{U_L} + \theta_{D_L}) + N_f (\theta_{\nu_L} +
\theta_{e_L}) + 2(\theta_h + \theta_{\chi^\pm}) - 2\theta_{gh}  \Big) \\ 
& = & \frac{11}{6} g_2^2 T^2~.
\eeqa{eq:thermgaugemassw0}
As in the previous case, any new particle in generic representation $r$ of the SM gauge group contributes to this expression
\beq
\Delta \Pi_{W_L^3}^{fermion,\ r} = \frac{g_2^2 T^2}{6} {\rm Tr}\ \big( T^3(r) T^3(r) \big), \ \ \ \ 
\Delta \Pi_{W_L^3}^{scalar,\ r} = \frac{g_2^2 T^2}{6} 2 {\rm Tr}\ \big( T^3(r) T^3(r) \big)~. 
\eeq{eq:w3arb}
In the particular case of an extra doublet we get 
\beq
\Delta \Pi_{W_L^3}^{scalar,\ 2} = \frac{g_2^2 T^2}{24} \ 2N_c (\theta_u + \theta_d), \ \ \ \ \Delta \Pi_{W_L^\pm}^{fermion,\ 2} = 
\frac{g_2^2 T^2}{24} \
 N_c (\theta_u + \theta_d)~.  
\eeq{eq:scal2piw3}

Now we calculate the thermal mass of the gauge boson $B$. Note that hereafter we use the $U(1)_Y$ gauge coupling $g'$ in the 
\emph{regular SM conventions}, and not in the SUSY unified conventions. In the SM we get 
\beqa \nonumber 
\Pi_{B_L} & = &\frac{{g'}^2 T^2}{216} \Big( 9N_f (\theta_{\nu_L} + \theta_{e_L} + 4 \theta_{e_R})  + 
18 (\theta_{\chi^\pm} + \theta_h) 
+ N_c N_f (\theta_{U_L} + \theta_{D_L} + 16 \theta_{U_R} + 4 \theta_{D_R})\Big) \\
& = & \frac{11}{6} {g'}^2 T^2~.
\eeqa{eq:bsm}
Any new BSM particle of hyper charge $Y$ contributes to this quantity as
\beq
\Delta \Pi_{B_L}^{fermion,\ Y} = N \frac{g^{\prime 2} T^2}{6} Y^2; \ \ \ \ 
\Delta \Pi_{B_L}^{scalar,\ Y} = N \frac{g^{\prime 2} T^2}{6} 2 Y^2~. 
\eeq{eq:blarb}
where the normalization in the first formula corresponds to a single Weyl fermion, and $N$ stands for the total number of complex degrees of freedom. For example, for a weak-singlet scalar with $N_c$ colors, $N_f$ flavors, and no exotic non-SM quantum numbers, we would simply have $N = N_c N_f$. 

Now we switch to the Higgs thermal mass. Pure gauge contribution to a scalar, charged under $SU(N)$ or $U(1)$ reads
\beq
\Pi_h^{SU(N)} = \frac{T^2 g_N^2}{4} C_N(r); \ \ \ \ \Pi_h^{U(1)} = \frac{T^2 {g'}^2}{4} Y^2
\eeq{eq:gauge2scal} 
where $r$ stands for the representation of the scalar, and $(T^a(r) T^a(r))_{ij} \equiv C_N(r) \delta_{ij}$. 
Note that the Higgs is a doublet of $SU(2)$ and  $C_2(r=2) = 3/4$. Hence, the gauge contribution in the SM reads
\beq
\Pi^{gauge}_h= \frac{3}{16} g_2^2 T^2 + \frac{1}{16} {g'}^2 T^2~.
\eeq{eq:higgsgauge}
Since in this paper we introduce different scalars with exotic representations, which theoretically can affect the EWPT, 
we list here for completeness quadratic Casimirs of the lowest representations of $SU(2)$ and $SU(3)$, namely 
for $SU(2)$: $C_2(2) = 3/4$, $C_2(3) = 2$, and for 
$SU(3)$: $C_3(3) = 4/3$, $C_3(6) = 10/3$ and $C_3(8) = 3$. 

To work out the contribution to the Higgs thermal mass from its self-couplings and Goldstone modes, we plug Eq.~\leqn{eq:higgsparametrization} into the thermal potential  and expand to the leading order in $T$. This yields
\beq
\Pi_h^{self} =  \Pi_{\chi_i}^{self} = \frac{\lambda T^2}{2}~. 
\eeq{eq:higgs}
Finally, it is necessary to take into account the contribution from the top quark. In our normalization, the Higgs coupling to quarks reads
\beq
\cL = \frac{y_t}{\sqrt{2}} h Q t^c~,
\eeq{eq:htoquraks}
so that the top mass is given by $m_t(\varphi) = y_t \varphi/ \sqrt{2} $. This leads to $\Pi_h^{top} = \frac{y_t^2 T^2}{4}$. Collecting all these contributions together
we get the thermal Higgs mass in the SM:
\beq
\Pi_h = \frac{3}{16} g_2^2 T^2 + \frac{1}{16} {g'}^2 T^2 + \frac{\lambda T^2}{2} + \frac{y_t^2 T^2}{4}.
\eeq{eq:pihall}
Note that in our conventions $y_t \approx 1$ and $v = 246$~GeV. 

Now we add to these calculations a BSM scalar in an arbitrary representation of the SM gauge groups, with the potential~\eqref{Vphi}. 
Its contribution to the thermal mass of the Higgs reads
\beq
\Pi_h^{NP} = N \frac{\kappa T^2}{12}~,
\eeq{eq:new2h}
where $N$ again stands for the total number of complex degrees of freedom. The self-contribution of the scalar with $N$ complex degrees of freedom reads
\beq
\Pi_\phi^{self} = \frac{\eta T^2}{6} (N + 1).
\eeq{selfie}
This expression of course agrees with~\eqref{eq:pihall} for $\eta \to \lambda$ and $N = N_c = 2$, as for the SM Higgs.